\documentclass[%
 reprint,
amsmath,amssymb,
aps,
pra,
floatfix,
]{revtex4-2}

\usepackage{booktabs}
\usepackage{graphicx}\usepackage[utf8]{inputenc}
\usepackage{xcolor}

\usepackage[caption=false]{subfig}
\usepackage{nicefrac}
\usepackage{mathtools}
\usepackage{centernot}

\usepackage{siunitx}
\usepackage{tabularx}

\usepackage{xr} %

\usepackage{natbib}

\usepackage{xurl}
\usepackage{hyperref} 
\hypersetup{urlbordercolor={1 1 1}}

\newcommand{\papertitle}{Nodal Braess's paradox and inertia destabilization with dynamic node and line failures in power grids}

\newcommand{\paperauthors}{Nubius Brandner$^{1,2,3,*}$, Frank Hellmann$^{1,*}$, Hans Würfel$^{1}$, Jürgen Kurths$^{1,4}$, Anton Plietzsch$^{5}$, Anna Büttner$^{1}$}
\newcommand{\paperaffiliations}{%
$^{1}$Research Domain Complexity Science, Potsdam Institute for Climate Impact Research (PIK), Potsdam, Germany.\\
$^{2}$Reiner Lemoine Kolleg c/o Reiner Lemoine Institut gGmbH, Berlin, Germany.\\
$^{3}$Institute of Theoretical Physics, Technische Universität Berlin, Berlin, Germany.\\
$^{4}$Department of Physics, Humboldt-Universität zu Berlin, Berlin, Germany.\\
$^{5}$Fraunhofer Research Institution for Energy Infrastructures and Geotechnologies IEG, Cottbus, Germany.\\
*Corresponding authors: brandner@pik-potsdam.de; hellmann@pik-potsdam.de
}

\usepackage{float}
\makeatletter
\let\origfigure\figure
\let\endorigfigure\endfigure
\newcommand{\forcesuppfigureshere}{%
  \renewenvironment{figure}[1][]{\origfigure[H]}{\endorigfigure}%
}
\makeatother

\newcommand{\beginsupplement}{
  \setcounter{section}{0}
  \setcounter{figure}{0}
  \setcounter{table}{0}
  \setcounter{equation}{0}

  \renewcommand{\thesection}{S\arabic{section}}
  \renewcommand{\thefigure}{\arabic{figure}}
  \renewcommand{\thetable}{S\arabic{table}}
  \renewcommand{\theequation}{S\arabic{equation}}

  \def\figurename{Supplementary Figure}

  \renewcommand{\theHsection}{S\arabic{section}}
  \renewcommand{\theHfigure}{S\arabic{figure}}
  \renewcommand{\theHtable}{S\arabic{table}}
  \renewcommand{\theHequation}{S\arabic{equation}}  

  \setcounter{enumi}{0}
  \setcounter{enumii}{0}
  \setcounter{enumiii}{0}
  \setcounter{enumiv}{0}

  \forcesuppfigureshere
} 

\begin{document}

\title{\papertitle}
\author{\paperauthors}
\affiliation{\paperaffiliations}
\date{\today}

\begin{abstract} %
\noindent
Large-scale power outages are typically caused by cascading failures. These unfold dynamically through complex interactions between network dynamics and individual component failures. In contrast, the study of cascading failures in physics has focused on analyzing line overloads in the quasi-static regime.
We introduce a new model that integrates the dynamics of node and line failures with a paradigmatic oscillator model for power grid synchronization. This enables us to investigate the collective cascading behavior of coupled failures for the first time. We study the impact of nodal robustness, the ability of nodes to tolerate transient disturbances, and inertia, the ability of nodes to resist frequency deviations, on cascade sizes. We discover two novel mechanisms driving system fragility:
i) While low inertia is widely considered a major challenge for power grids, we find that high inertia can amplify cascade sizes unless accompanied by appropriate adjustments of other dynamical properties. ii) Further, we find that an increase in the robustness of individual nodes can paradoxically lead to larger cascades. This latter phenomenon constitutes a novel type of Braess's paradox. Understanding such counterintuitive collective effects may become central for achieving resilient future power grids.

\end{abstract}

\maketitle

\section{Introduction} %

Large-scale power grid blackouts are almost always caused by cascading failures~\cite{unionfortheco-ordinationoftransmissionofelectricityFinalReportSystem2007,departmentforbusinessenergy&industrialstrategyGBPowerSystem2020,bialekWhatDoesPower2020}. With the continued electrification of transportation and heating, the consequences of such disruptions are expected to become even more severe.
Understanding cascading failures is thus a central challenge in network science, statistical physics, and engineering~\cite{motterCascadebasedAttacksComplex2002,crucittiModelCascadingFailures2004,artimeRobustnessResilienceComplex2024,schaferDynamicallyInducedCascading2018,sturmerIncreasingResilienceTexas2024,zakariyaSystematicReviewCascading2023}.
Failure cascades have a rich phenomenology that links a wide range of mathematical and theoretical approaches, while also being of immense practical relevance.

At the intersection of power grids and statistical physics, failure cascades have typically been studied as sequences of steady states~\cite{sturmerIncreasingResilienceTexas2024,kinneyModelingCascadingFailures2005,pahwaAbruptnessCascadeFailures2014,plietzschLocalVsGlobal2016,rohdenCascadingFailuresAc2016}, with a focus on line failures.
However, real-world cascades in power grids inherently evolve dynamically. Failures during blackouts are governed not only by network topology and static power flow distribution, but also by the collective dynamics of the entire system during transients. Additionally, real power grids also show node failures, such as generator trips caused by frequency deviations, that interact dynamically with line overloads. In this paper, we study collective phenomena of systems of dynamically coupled node and line failures for the first time systematically. Our nodes aggregate local generation and consumption units, where some of them can fail due to grid conditions. We uncover and explain two counterintuitive phenomena: More robust nodes and higher inertia cause larger cascades.

Inertia, which determines the rate at which the frequency throughout the grid changes when a power imbalance arises, is a central factor for the stability of a power grid. Low system inertia is regarded as a central stability challenge for power grids dominated by renewable energy sources, as solar and wind generation do not inherently provide inertia~\cite{ulbigImpactLowRotational2014, milanoFoundationsChallengesLowInertia2018, hartmannEffectsDecreasingSynchronous2019, rydingorjaoOpenDatabaseAnalysis2020, nnoliSpreadingDisturbancesRealistic2021, entso-eFrequencyStabilityLongterm2021}. 
This requires the large-scale deployment of new technologies, in particular inverters with appropriate control strategies, and the procurement of inertia through market mechanisms to compensate for the reduction. Previous studies have shown that inertia reduces frequency fluctuations and helps prevent generator failures~\cite{ulbigImpactLowRotational2014, poollaOptimalPlacementVirtual2017, tamrakarVirtualInertiaCurrent2017, jacquodOptimalPlacementInertia2019, pagnierInertiaLocationSlow2019, fritzschStabilizingLargeScaleElectric2024, parkOptimalLocationReinforced2025}.

Detailed engineering models of individual systems allow the investigation of individual cascade events in great detail~\cite{zakariyaSystematicReviewCascading2023, wangDynamicCascadingFailure2021}. However, their high computational cost and limited generalizability restrict their broader applicability. Moreover, dynamical models for the devices that will dominate future power grids, such as grid-following or grid-forming inverters, are still under active development, and there are no established consensus models. The large-scale collective dynamics of node failures coupled with line failures have not been extensively investigated in this context.

However, in the past, more conceptual dynamical models have been able to reveal novel phenomena and vulnerabilities that arise from the system's collective dynamics without using highly complicated engineering models ~\cite{simonsenTransientDynamicsIncreasing2008,yangCascadingFailuresContinuous2017,frascaControlCascadingFailures2021,galindo-gonzalezDecreasedResiliencePower2020,olmiMultilayerControlSynchronization2024}. Schäfer et al.~\cite{schaferDynamicallyInducedCascading2018} have shown that a dynamic transmission line failure model indicates a significant increase in a network's vulnerability compared to its static counterpart. For a case-study of dynamic load-shedding avalanches, \cite{parkOptimalLocationReinforced2025}~noted that inertia can have a counterintuitive impact in some scenarios.

In this work, we introduce a dynamical model that contains both node and line failures. 
By integrating node failures and their dynamic interaction with line failures, we introduce a framework that couples frequency-driven node disconnections with overload-induced line outages. This approach captures key characteristics of real-world phenomena in power grids from a statistical physics perspective that have not been studied yet.

Using this model, we discover two qualitatively novel phenomena. In contrast to the established notion that high inertia stabilizes a system, our results show that when considering dynamical line failures, high inertia can increase a network's vulnerability to cascades.
This can be mitigated by adjusting further dynamic parameters on the nodes. Considering the collective dynamics of the grid adds considerable nuance to the established inertia narrative.

Paradoxically, we also find that increasing nodal robustness can reduce overall network stability.
This behavior is reminiscent of Braess's paradox, a counterintuitive phenomenon originally found in traffic networks~\cite{braessUeberParadoxonAus1968} and later in power grids~\cite{schaferUnderstandingBraessParadox2022,witthautBraesssParadoxOscillator2012,colettaLinearStabilityBraess2016,tchuisseuCuringBraessParadox2018a}, where increasing the capacity of existing lines or adding new lines may reduce the overall system capacity.
Our work extends Braess's paradox to the domain of dynamic cascading failures, discovering a phenomenon of genuinely new nature.
This paradoxical behavior is not confined to individual scenarios and dominates the expected behavior in our setting. We identify two underlying mechanisms: one arising from the statistical properties of the power flow before the cascade and one from the local equilibration dynamics near the location of disruptions.

These findings may have important implications for the design of future grid codes, which are currently being revised in response to the transition towards renewable energy sources~\cite{entso-esystemdevelopmentcommitteePositionNeedNational2025, entso-eRDIRoadmap202420342024, internationalrenewableenergyagencyGridCodesRenewable2022}. These results also highlight the need to evaluate the systemic impact of individual component dynamics on collective system dynamics.

In this study, we first introduce a dynamical framework that couples node and line failures. We analyze the impact of inertia and describe the underlying mechanisms for the loss of stability in high-inertia scenarios. We then demonstrate that a fully analytically tractable conceptual model allows us to identify further dynamical adjustments that mitigate the effect. Finally, we provide evidence for a dynamical Braess's paradox and analyze the mechanisms behind it.

\section{Results} %

\subsection{A dynamic cascade model}  \label{sec:model} %

We first provide the network dynamics model in which we study node as well as line failures. We then introduce our failure mechanisms and explain the experimental set-up for studying their interactions in cascades.

\subsubsection{Network dynamics} \label{sec:coupling_node_line} %

As a network dynamics model, we use the structure preserving Bergen-Hill model for power grids~\cite{bergenStructurePreservingModel1981}. Crucially for our purposes, it has two different types of dynamical nodes, which enabled us to implement dynamic node failures.

In a power grid, nodes typically aggregate several local generation and consumption units.
The net power injection $P_k$ at node $k$ is the sum of these local contributions. For $P_k > 0$, the supply is larger than the demand and node $k$ operates as an effective generator, injecting power into the grid. For $P_k < 0$, it acts as an effective power consumer.

The central dynamical distinction is not between loads and generators, but between units that provide inertia and those that do not.
In traditional power grids, only conventional generators provided inertia, whereas future grids will utilize inverters that connect renewable energy sources, and batteries or controllable loads to provide a programmable inertial response. For simplicity, we refer to them as grid-forming units.
To distinguish between power injections from units with and without inertial response, we decompose the power injection into $P_k = P_k^{\text{GFM}} + P_k^{\text{GFL}}$, with a grid-forming part  $P_k^{\text{GFM}}$ that has inertia and a grid-following part $P_k^{\text{GFL}}$ that does not.

In the Bergen-Hill model, nodes that contain grid-forming units are modeled by the widely used swing equation~\cite{machowskiPowerSystemDynamics2020, nishikawaComparativeAnalysisExisting2015, guoOverviewsApplicationsKuramoto2021} (also referred to as the second-order Kuramoto model), which describes the key dynamic synchronization behavior of power grids:
\begin{align}
  \label{eq:swing_equation_a}
  \frac{d}{dt}\theta_k & = \omega_k,\\
  \label{eq:swing_equation_b}
  I_k \frac{d}{dt}\omega_k & = P_k - D \omega_k - \sum_{m=1}^{\mathrm{N}} x_{km}^{-1} \sin (\theta_k - \theta_m).
\end{align}
For a network consisting of $N$ nodes, each node $k$ represents a device with effective inertia $I_k$ and damping $D$. The reactance $x_{km}$ sets the coupling strength between individual nodes. Each node is described by its voltage phase angle $\theta_k$ and its angular frequency $\omega_k$ relative to a reference frame of $2\pi(50 \textrm{ or } 60)$ Hz. 

Grid-following nodes lacking inertia are modeled in the Bergen-Hill model by first-order Kuramoto oscillators:
\begin{equation} \label{eq:bergen_hill}
   \tau \dot{\theta_k} = P_k^{\text{GFL}} - \sum_{m=1}^{\mathrm{N}} x_{km}^{-1} \sin (\theta_k - \theta_m),
\end{equation}
with time constant $\tau$. We assume that damping $D$ and the constant $\tau$ are homogeneous throughout the network.

\subsubsection{Coupling dynamic node and line failures} \label{sec:cascading_failure_model}

Failure cascades occur when the failure of one component causes a shutdown of a second component. In power grids, this shutdown is typically not caused by physical damage to the second component. Instead, protection mechanisms disconnect the component when predefined operational limits are exceeded.

To study the interplay of node and line failures, we introduce a novel dynamic node failure model. Although these node and line failure models are not fully representative of protection mechanisms in real power grids, they are suitable for the level of abstraction and time-scale in this study. As in the model by Schäfer et al.~\cite{schaferDynamicallyInducedCascading2018}, we choose a fully dynamic model of failures. Rather than waiting until the system reaches a new steady state before evaluating the operational boundaries~\cite{witthautNonlocalEffectsCountermeasures2015, plietzschLocalVsGlobal2016, rohdenCascadingFailuresAc2016}, the protection mechanisms are in effect throughout the dynamical simulation. Whenever a failure occurs during the transient response (dynamical response of system variables), the network is modified instantaneously and the simulation proceeds on this modified system.

\paragraph{Node Failure Model:}
Components in a power grid are designed to operate within specific frequency bounds~\cite{bialekWhatDoesPower2020, blumeElectricPowerSystem2017} and are disconnected when these bounds are exceeded.
We model this protection mechanism by failing the grid-forming units at a swing equation node when the angular frequency deviation becomes too large. This is quantified by a bound $f_b$ on the frequency $f_k = \omega_k / (2 \pi)$. Failure is triggered by the condition:
\begin{equation} %
\label{eq:node_failure_condition}
|f_k(t)| > f_b.
\end{equation}
When this condition is met, the grid-forming power injection $P_k^{\text{GFM}}$ of the affected node is set to zero, and the node transitions from a second-order swing equation node described by Eq. (\ref{eq:swing_equation_b}) to a first-order node governed by Eq. (\ref{eq:bergen_hill}).
This transition models the three following features:
\begin{enumerate}
    \item Disconnection takes place much faster than synchronization.
    \item The power injection that belongs to the grid-forming contribution is set to zero.
    \item The node no longer contributes inertia.
\end{enumerate}
It is important to note that the network topology is preserved after the failure, which means that power can still flow through the affected node. We do not consider further nodal failures and nodal protection mechanisms. 
This model does not aspire to be a complete model of all nodal failures relevant in real blackouts. Most importantly, as we do not consider voltage amplitude dynamics, we do not capture failures due to over/under-voltages, and we do not consider load-shedding.

\paragraph{Line Failure Model:}

Protection settings for transmission lines in power grids are determined mainly by ohmic heating limits, and typically indicated in terms of the current on the line. When these limits are exceeded, the affected line is automatically disconnected to prevent damage.
We model line failures by deactivating the corresponding line in the graph.

In the Bergen-Hill model, which does not integrate voltage fluctuations, current limits can be equivalently expressed as limits on the apparent power $S$. This includes both the active power flowing on the line, and the reactive power, which measures the amount of energy stored in the line itself. It is discussed in detail in Methods~\ref{sec:apparent_power}, in Eq. \eqref{eq:maximum_apparent_power}. In our setting $S$ equates to
\begin{equation}
|S_{km}|= \frac{2}{x_{km}}\left|\sin\left(\frac{\theta_k - \theta_m}{2}\right)\right|\; .
\end{equation}

We define the line failure condition as:
\begin{equation}
\label{eq:line_failure_condition}
|S_{km}(t)|  > C_{km} =  \alpha x_{km}^{-1},
\end{equation}
with capacity $C_{km}$ and safety parameter $\alpha$. Typical values for power grids are $\alpha \approx 0.7$~\cite{gonenElectricalPowerTransmission2014}, which will be used throughout this work. When the line failure condition is met, we set $x_{km}^{-1}$ to zero. Note that the capacity $C_{km}$ is an absolute capacity that is independent of the system's initial state.
This line failure model is similar to the one studied in~\cite{schaferDynamicallyInducedCascading2018}, where active rather than apparent power was used. %

\subsubsection{Cascade experiments} \label{sec:WS_networks}

To study the combined effect of node and line failures systematically, we conducted experiments on failure cascades induced by the initial failure of a single line.

In each simulation, we initialize the system at a stable operating state, where all nodes are in synchrony at the reference angular frequency, i.e. $\omega_k = 0 \ \forall k \in \{1, \ldots, N\}$, implying $\sum_k P_k = 0$. 
We then remove a line $i$, and simulate the system's response, including potential further failures, until a final steady state is reached in which the failure conditions in Eqs. (\ref{eq:node_failure_condition}) and (\ref{eq:line_failure_condition}) are not met. In the final state we evaluate
\begin{align}
    &n_i &&\text{ the number of node failures,}\\
    &l_i &&\text{ the number of line failures,}\\
    F_i &= n_i + l_i &&\text{ the total number of failures.} 
\end{align}

We simulate all possible triggering lines $i$, and evaluate the mean number of failures over all lines in the network as $\langle n\rangle$, $\langle l\rangle$ and $\langle F\rangle$.

We apply the cascading failure model to two types of networks: a realistic power grid test system and an ensemble of generic synthetic networks.
For the first case, we used the RTS-GMLC power grid test system~\cite{barrowsIEEEReliabilityTest2020}, which represents the topology and parameters of a transmission network, and 
consists of $N=73$ nodes and $M=108$ transmission lines.
The generator nodes are modeled by the swing equation, while load nodes follow the first order model, see Eqs. (\ref{eq:swing_equation_b}) and (\ref{eq:bergen_hill}). The test system does not specify values for the damping parameter $D$ and the time constant $\tau$. We therefore fix a homogeneous damping of $D = \SI{0.1}{\second}$ and set the time constant to $\tau = \SI{0.01}{\second}$, ensuring that the load dynamics evolve significantly faster than those of the generator nodes, in accordance with the original model assumptions of~\cite{bergenStructurePreservingModel1981}.

To investigate general mechanisms driving the cascades in depth and look beyond the concrete power grid case, we also study an ensemble of 32 networks generated by the Watts–Strogatz (WS) model~\cite{wattsCollectiveDynamicsSmallworld1998}, a fundamental random graph generation model.
We construct networks with $N = 100$ nodes and $M=200$ lines, with a mean degree of $k = 4$, initially connecting each node to four neighbors. As we have $k = 4$, the network contains no dead ends~\cite{menckHowDeadEnds2014}. We further set the rewiring probability to $\beta=0.5$, see Methods~\ref{sec:methods_WS_choice}.
 
As opposed to the power grid test case, all nodes are modeled by swing equations in the initial steady state, see Eq. (\ref{eq:swing_equation_b}).
We set the damping of the swing equation nodes and the time constant of the first-order nodes to $D = \tau = \SI{1}{\second}$. We further set the inverse reactance to $x_{km}^{-1} = 3$, as in~\cite{buttnerAmbientForcingSampling2022}.
The power injections $P_k^{\text{GFM}}$ and $P_k^{\text{GFL}}$ are drawn from a normal distribution with mean $\mu = 0$ and variance $\sigma^2 = 1$. %
To obtain an initially balanced system, we balance the total power injections, see Methods~\ref{sec:methods_balancing}.
This setup has the same dominant time-scale for first- and second-order dynamics ($\tau \approx I$), which is in contrast to the power grid test case, where $\tau \ll I$. It is not designed to reflect realistic power grids directly, but rather to gain maximum insights into novel collective phenomena for cascading failures.

In these experiments, we systematically varied the inertia $I_k$ and the frequency bounds $f_b$. For the WS networks, the inertia is homogeneous: $I_k = I$. For the power grid case, we scale the inertia homogeneously by a constant factor (see Methods~\ref{sec:methods_parameter_choice} for the specific parameters). Furthermore, we use the per-unit system~\cite{machowskiPowerSystemDynamics2020} by replacing the ``real''
parameters with dimensionless multiples with respect to reference values. Here, the power \mbox{$P_{\textrm{per unit}}=1$~p.u.} corresponds to $P_{\textrm{real}}=\SI{100}{\mega\watt}$.

\subsection{Inertia-induced instability through increased line failures} \label{sec:inertia_effect} %

As noted in the introduction, it is widely assumed that inertia stabilizes the power grid's dynamics. Our first major finding is that inertia can destabilize the models considered here. This is driven by the impact of inertia on line dynamics.

\begin{figure} %
  \includegraphics[width=0.49\linewidth]{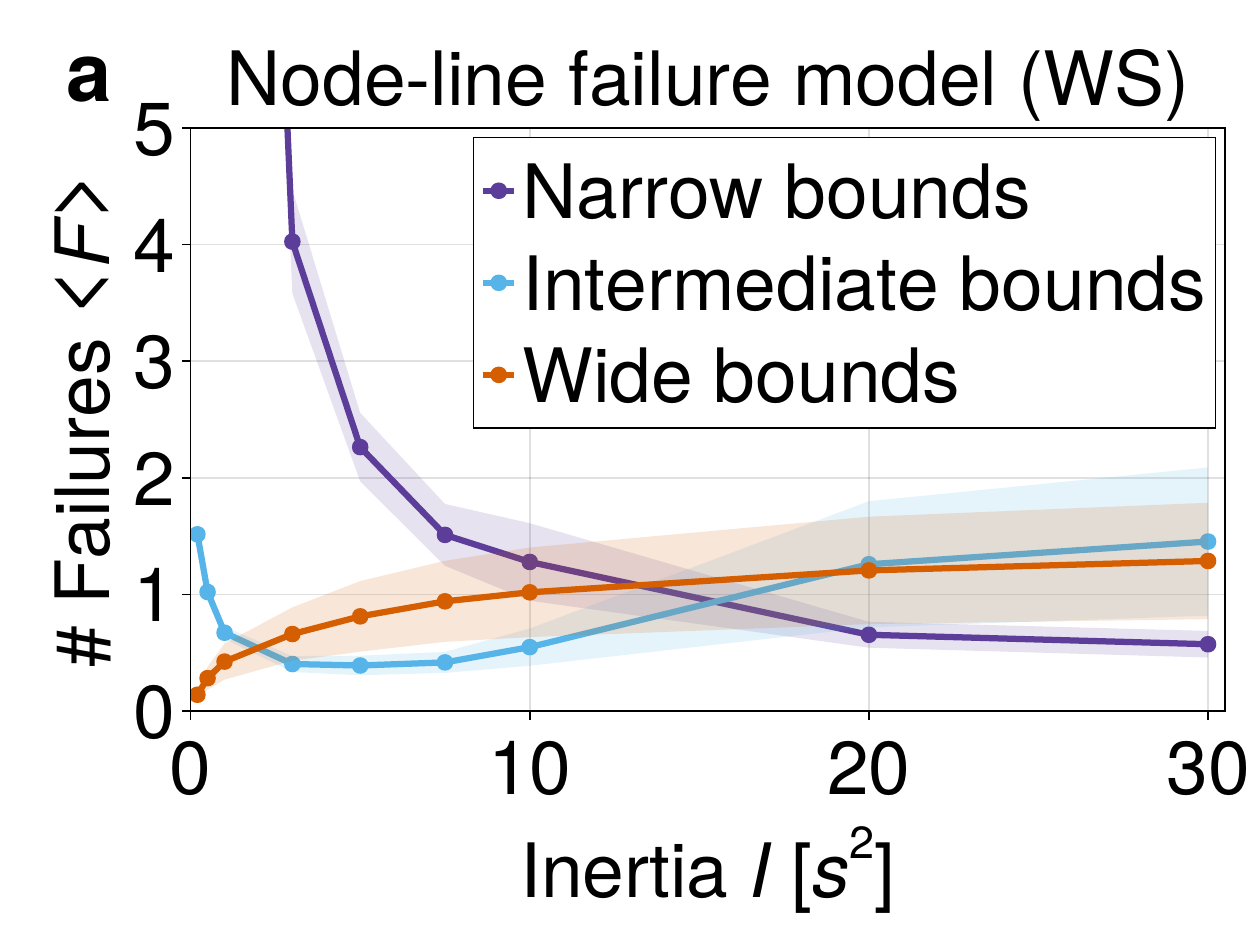}
  \includegraphics[width=0.49\linewidth]{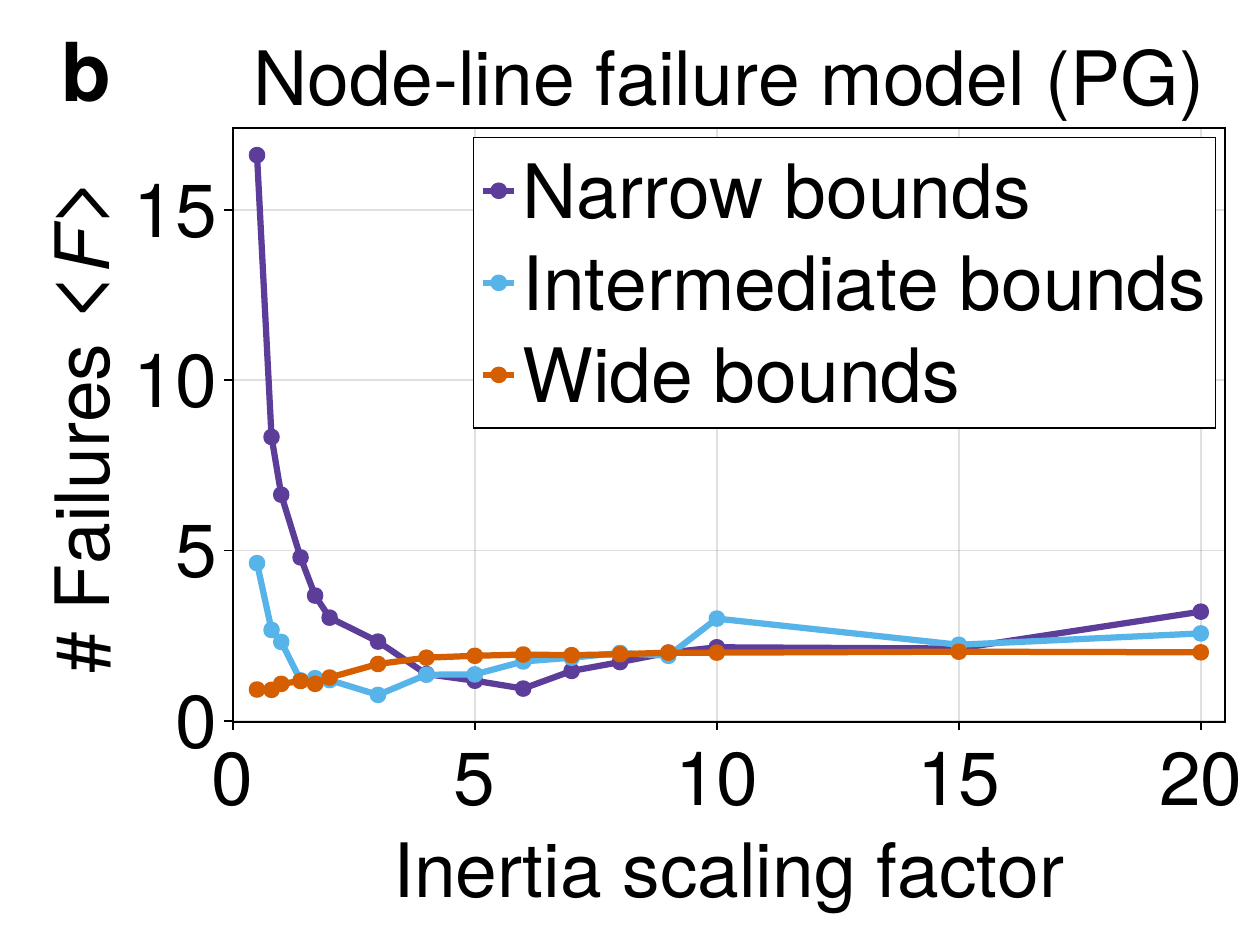}
  \\
  \includegraphics[width=0.49\linewidth]{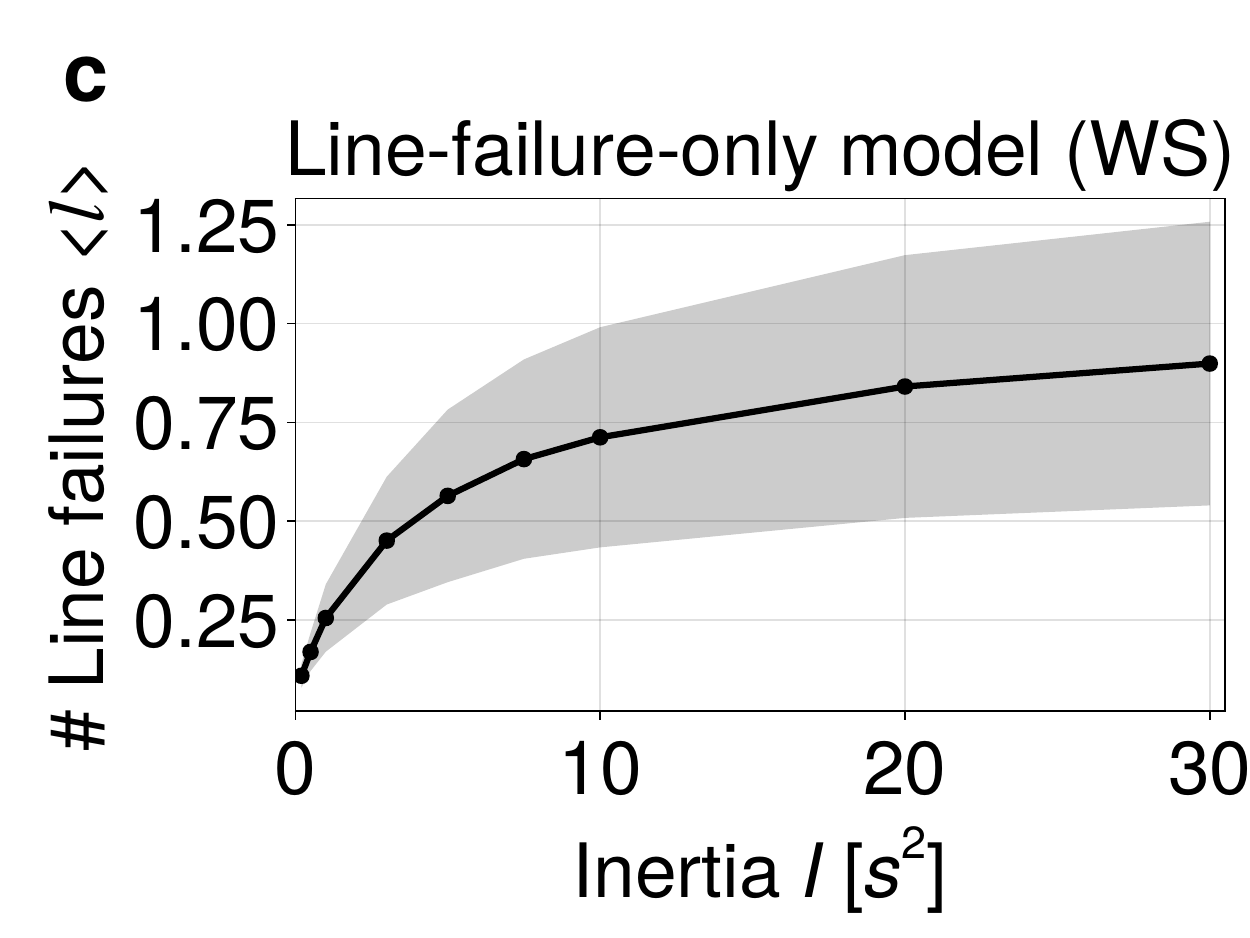}
  \includegraphics[width=0.49\linewidth]{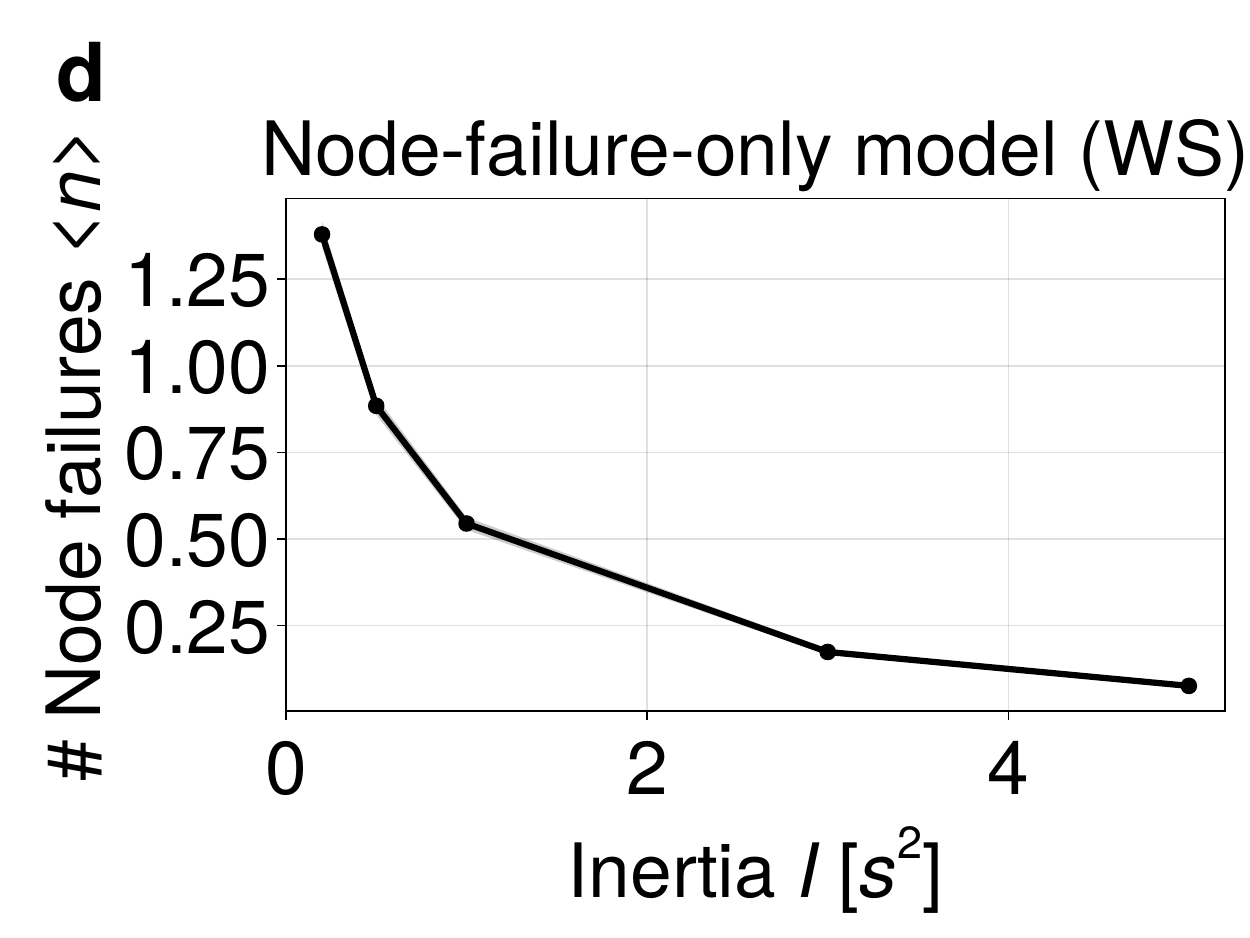}
  \\
  \includegraphics[width=0.49\linewidth]{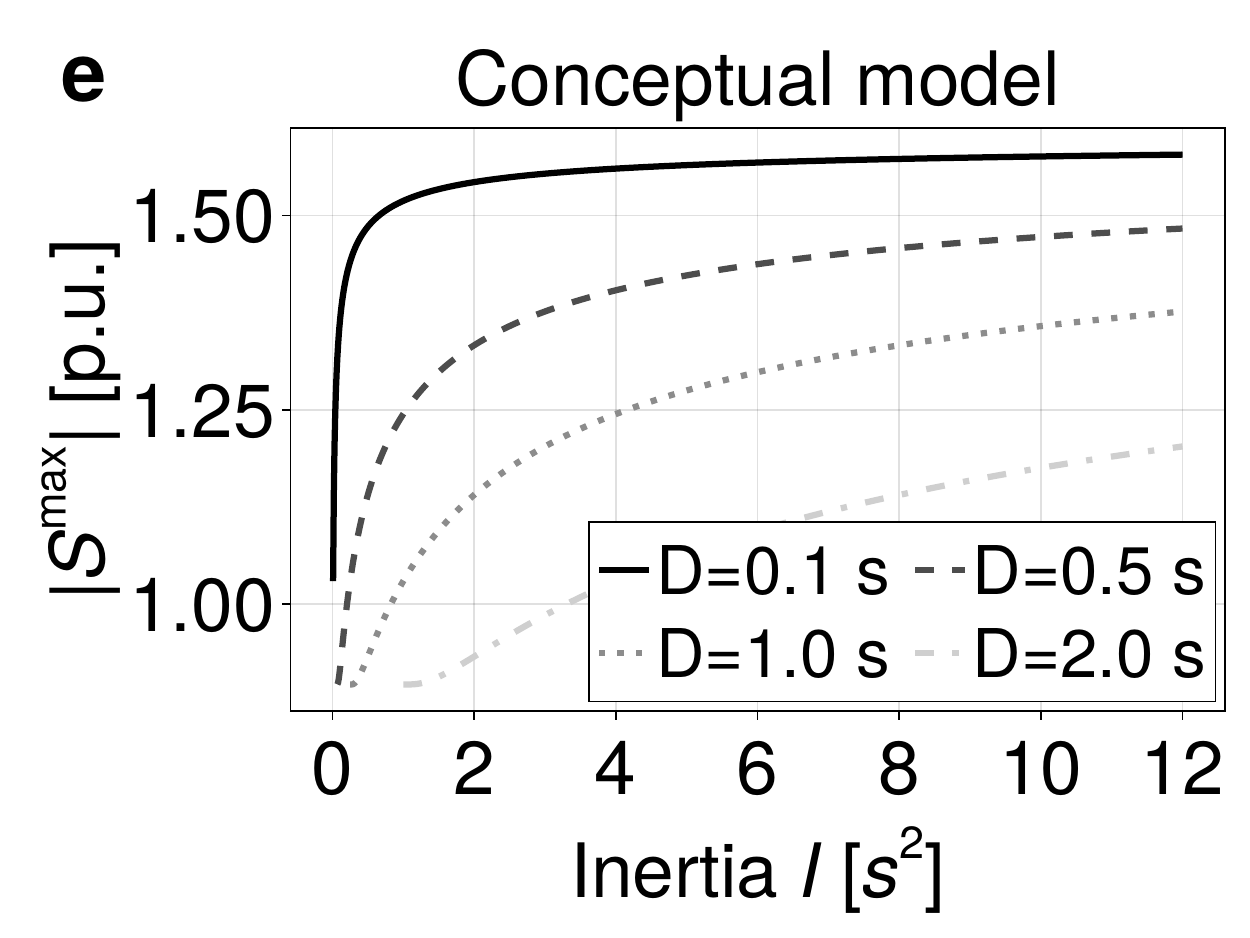}
  \includegraphics[width=0.49\linewidth]{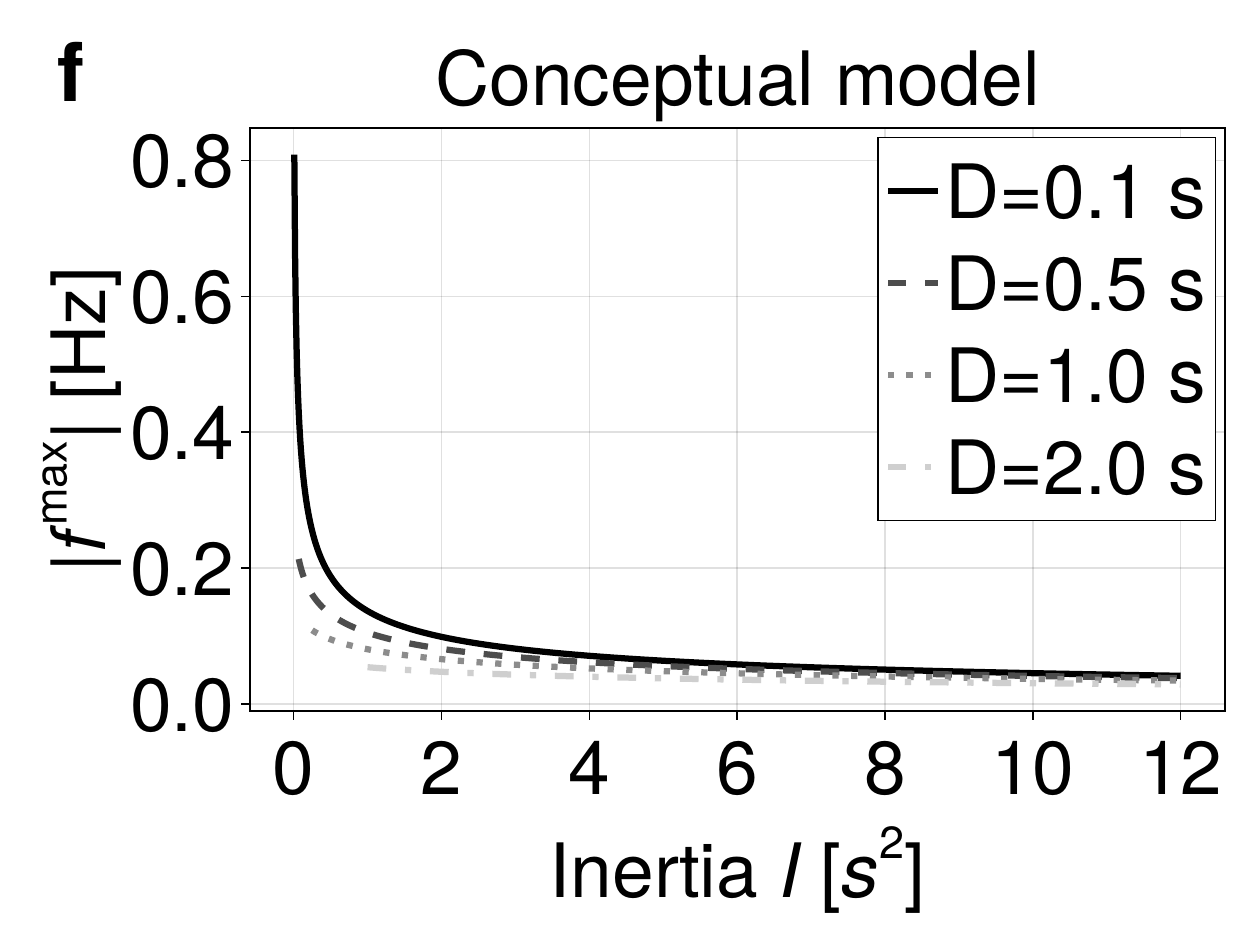}
  \caption{Network stability as a function of inertia $I$. Depending on the frequency bounds, inertia can destabilize power grids. An ensemble of 32 Watts–Strogatz (WS) networks with 100 nodes and 200 lines is analyzed in \textbf{a}, where the coupled node-line failure model allows for both node and line failures, in \textbf{c}, where the decoupled line-failure-only model allows exclusively for line failures and in \textbf{d}, where the decoupled node-failure-only model allows exclusively for node failures. The standard error is shown in shaded bands. 
  For narrow bounds, the summed ensemble average of line and node failures $\langle F\rangle$ in \textbf{a} scales similarly to the node failures $\langle n\rangle$ in \textbf{d}. For wide bounds, $\langle F\rangle$ in \textbf{a} scales similarly to the line failures $\langle l\rangle$ in \textbf{c}. At intermediate bounds, there is an optimal, intermediate inertia value for $\langle F\rangle$.
  The RTS-GMLC power grid (PG) test system~\cite{barrowsIEEEReliabilityTest2020} demonstrates a similar behavior in \textbf{b}. Here, $\langle F\rangle$ is the average of all lines of a single network, and the inertia parameters at the nodes are heterogeneous and are thus scaled by a homogeneous factor.
  The behavior in \textbf{c} and \textbf{d} can be understood by a basic conceptual model, in which a single node is connected by a single line to the rest of the network which remains approximately static. 
  The analytical solution shows that when inertia $I$ varies, the maximum line loading $|S^{\text{max}}|$ behaves qualitatively like the line failures $\langle l\rangle$ in the line failure model \textbf{c}, whereas the maximum frequency $|f^{\text{max}}|$ behaves similar to the number of node failures $\langle n\rangle$ in the node failure model \textbf{d}.
  Furthermore, in \textbf{e} and \textbf{f} both $|S^{\mathrm{max}}|$ and $|f^{\mathrm{max}}|$ decrease as damping $D$ increases.} 
  \label{fig:inertia_effect} %
\end{figure}

Figures~\ref{fig:inertia_effect}a and \ref{fig:inertia_effect}b show how the expected number of failures $\langle F\rangle$, depends on inertia for the WS ensemble and the power grid, respectively.
We consider three regimes of nodal robustness, defined by the nodes having narrow, intermediate, and wide frequency bounds $f_b$ (see Eq. (\ref{eq:node_failure_condition})). We expect nodal failures to drive the cascade dynamics for narrow bounds, while line failures should dominate for robust nodes with wide bounds.
In both systems, we find that inertia can be destabilizing on the grid for both intermediate and wide bounds. For intermediate bounds, the dependence is not monotonic but an optimal level of inertia exists. See Supplementary Fig.~\ref{fig:WS_coupled_model_node_and_line_failures_separately} for low inertia values at narrow bounds for the WS networks.

In contrast, for narrow frequency bounds, where we expect the system to be driven by node failures, the system behavior agrees with expectations from the literature~\cite{ulbigImpactLowRotational2014}. The behavior at intermediate bounds appears to be a combination of that with narrow and wide bounds.

An example illustrating how varying inertia $I$ leads to different trajectories and cascading behaviors within a single network is given in Supplementary Fig.~\ref{fig:WS_trajectories_inertia_variation}.

\subsubsection{Mechanism of inertia-induced destabilization}
To understand the mechanism behind the inertia-induced destabilization,
we consider a basic conceptual model: a single node, connected by a single line to the rest of the network, with no power flowing on the connecting line in the initial state. 
If the power injected into the node jumps by a small amount $\Delta P$, and we assume that the rest of the system remains static, we can linearize the system dynamics and calculate the maximum line loading $|S^{\text{max}}|$ and maximum frequency $|f^{\text{max}}|$ during the transient, see Methods~\ref{sec:methods_analytic_model}:
\begin{equation} \label{eq:S^max}
  |S^{\max}| = 2 \left| \sin\!\left( \frac{\Delta P}{2} \left[ 1 + \exp\!\left( -\frac{\pi}{\sqrt{\eta - 1}} \right) \right] \right) \right|
\end{equation}
\begin{equation} \label{eq:f^max}
  |f^{\text{max}}| = \frac{|\Delta P|}{2\pi\sqrt{I}} \exp\!\left( -\,\frac{\arccos(\eta^{-1/2})}{\sqrt{\eta-1}}\right),
  \ \eta = \frac{4 I}{D^2}.
\end{equation}

Figs.~\ref{fig:inertia_effect}e and~\ref{fig:inertia_effect}f illustrate the dependence of maximal line loading and maximal frequency on the node parameters inertia $I$ and damping $D$.
We observe that the maximum line loading actually increases with inertia, while the maximum frequency deviation drops.

The conceptual model suggests that power imbalances lead to larger transient line loads and smaller transient frequency deviations throughout a cascade if inertia is high. We thus expect line failures to increase with inertia while node failures decrease. To verify that this is indeed the mechanism driving the inertia-induced fragility of the model, we analyze the model's variants, in which only node or line failures occur.

The results are shown in Figs.~\ref{fig:inertia_effect}c and~\ref{fig:inertia_effect}d.  In Fig.~\ref{fig:inertia_effect}c, only line failures are modeled (line-failure-only model), meaning that the line failure condition in Eq. (\ref{eq:line_failure_condition}) is monitored continuously, while the node failure condition in Eq. (\ref{eq:node_failure_condition}) is ignored (see Supplementary Fig.~\ref{fig:line_failures_test_case} for power grid), and vice versa for node failures (node-failure-only model) in Fig.~\ref{fig:inertia_effect}d.

Under variation of inertia, we find that $|S^{\text{max}}|$ in Fig.~\ref{fig:inertia_effect}e behaves the same way as the number of line failures $\langle l\rangle$ for the line-failure-only model in Fig.~\ref{fig:inertia_effect}c, and $|f^{\text{max}}|$ in Fig.~\ref{fig:inertia_effect}f behaves the same as the number of node failures  $\langle n\rangle$ in Fig.~\ref{fig:inertia_effect}d for the node-failure-only model.

This conceptual model appears to be highly predictive, and the qualitative agreement with the results for decoupled models is remarkable given the strongly simplifying assumptions in the conceptual model, see Methods~\ref{sec:methods_analytic_model}.

This mechanism is further confirmed by considering not just the total number of failures, but also $\langle n\rangle$ and $\langle l\rangle$ separately in the full model (Supplementary Fig.~\ref{fig:WS_coupled_model_node_and_line_failures_separately}).
For narrow bounds, line failures are virtually absent for all inertia values, while node failures follow the behavior of the node-failure-only model. For wide bounds, both line failures and node failures increase with inertia, and exhibit the same qualitative behavior as in the line-failure-only model, indicating that the observed node failures are caused by failures of adjacent lines. For intermediate bounds, node failures first decrease, while line failures increase as inertia is increased. Eventually, node failures start to increase again, proportionally to the increase in line failures. There is a trade-off between inertia, as it prevents node failures and increases line failures, and an optimal level of inertia exists.

\subsubsection{The role of Damping}

The analytic model in Eqs. \eqref{eq:S^max} and \eqref{eq:f^max} also predicts that both $|S^{\text{max}}|$ and $|f^{\text{max}}|$ decrease as damping $D$ increases. Further experiments confirm this expectation (Supplementary Fig.~\ref{fig:WS_vary_D_only_uebergang_lines+nodes}): the number of failures $\langle F\rangle$ decreases with $D$, independent of the frequency bounds $f_b$. The analytic model also suggests that jointly scaling inertia and damping can eliminate the problem of inertia-induced line overloads. Indeed, we find that keeping $\frac{D^2}{I}$ or $\frac{D}{I}$ constant addresses the issue (Supplementary Fig.~\ref{fig:WS_uebergang_lines+nodes_different_scalings}), where the latter performs slightly better. The subtle interplay of damping and inertia has been highlighted in different contexts before. For example, in~\cite{delgiudiceEffectsInertiaLoad2021} it was found that if the damping is too small, inertia becomes ineffective at limiting frequency excursions. This complementary analysis provides further confidence in the suggestion that damping and inertia should scale together.

\subsection{When nodal robustness destabilizes the grid: a dynamical Braess's paradox} \label{sec:breaessness_effect} %

Our second major finding is that increasing the frequency bounds $f_b$, and thus enhancing nodal robustness, can paradoxically increase the network's vulnerability to cascades, a nodal analog of Braess's paradox. This effect is already visible in Figs.~\ref{fig:inertia_effect}a and~\ref{fig:inertia_effect}b, where the curves corresponding to the widest frequency bounds, and thus to the highest nodal robustness, exhibit more failures than those with narrower bounds for a range of inertia values. We find that this effect persists very robustly across all inertia values (see Supplementary Fig.~\ref{fig:heatmap_f_b_vs_inertia}). It is not confined to individual trigger lines and failure scenarios, but rather represents the typical behavior. Next, we present detailed evidence supporting this effect, and then discuss a concrete example in which it occurs. Finally, we outline a detailed investigation of the underlying mechanisms.

\begin{figure}
  \includegraphics[width=\linewidth]{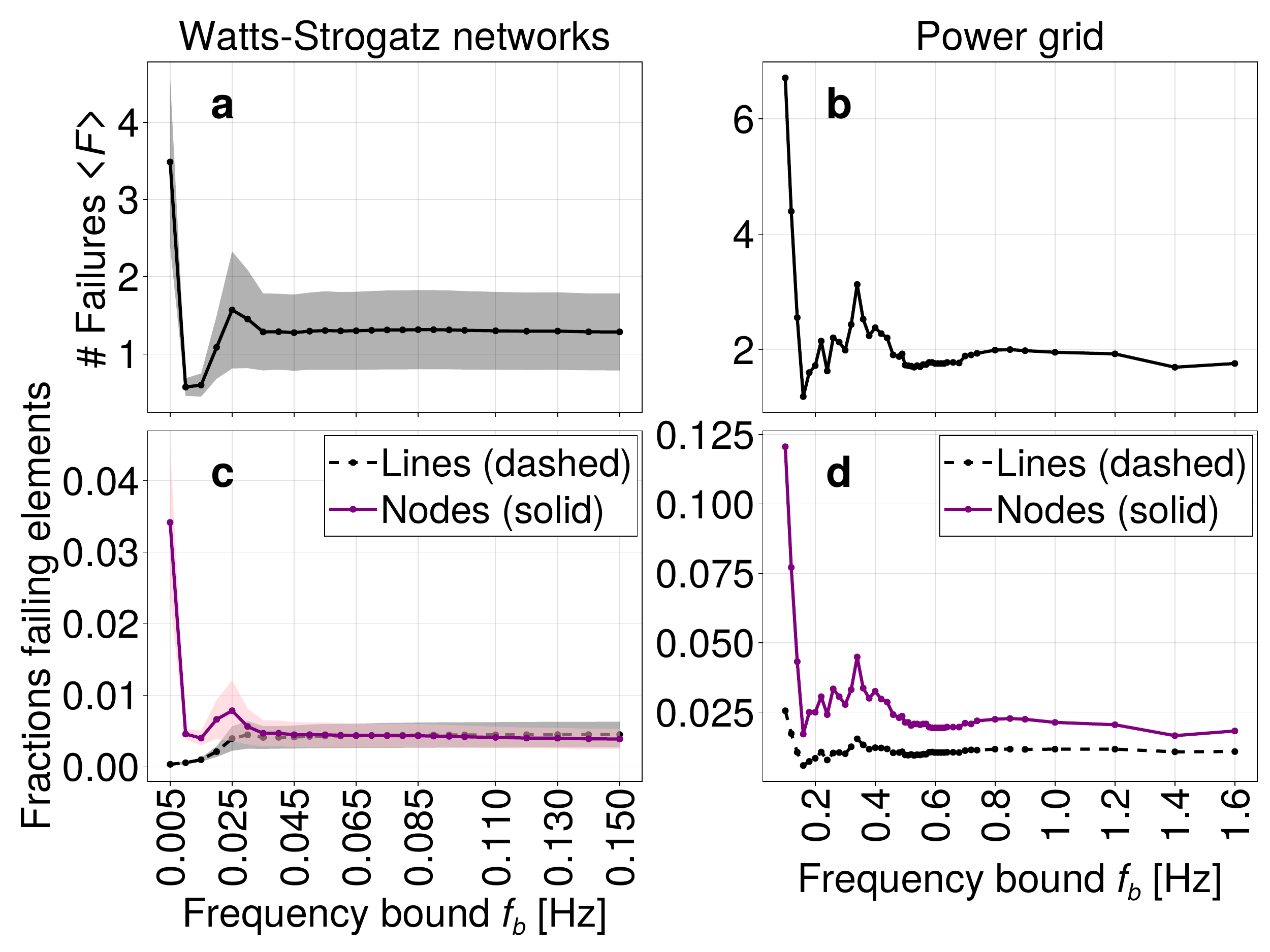}  
  \caption{Number of failures as a function of the frequency bounds $f_b$ of nodes. Panels \textbf{a},\textbf{c} show results for the Watts–Strogatz networks, and \textbf{b},\textbf{d} for the power grid test case. Panels \textbf{a},\textbf{b} show the summed average of all failures $\langle F\rangle$, while \textbf{c},\textbf{d} show the fractions of node failures $\frac{\langle n\rangle}{N}$ and line failures $\frac{\langle l\rangle}{M}$ separately. The standard error is shown in shaded bands. Widening the frequency bounds $f_b$ can paradoxically increase the number of failures, revealing a local maximum and indicating a dynamical analog of Braess's paradox.}
  \label{fig:f_b_vs_failures}
\end{figure}

Figure~\ref{fig:f_b_vs_failures} shows the number of failures as a function of the frequency bounds $f_b$. Increasing the frequency bounds $f_b$, and thereby increasing nodal robustness past a certain point increases the number of failures, and thus the grid's vulnerability to cascading failures.
This reduction in resilience is observed for both the WS ensemble Fig.~\ref{fig:f_b_vs_failures}a and the power grid test case Fig.~\ref{fig:f_b_vs_failures}b. Both feature local minima and maxima in the total number of failures~$\langle F\rangle$. %

\begin{figure}
  \center
  \includegraphics[width=0.5\linewidth]{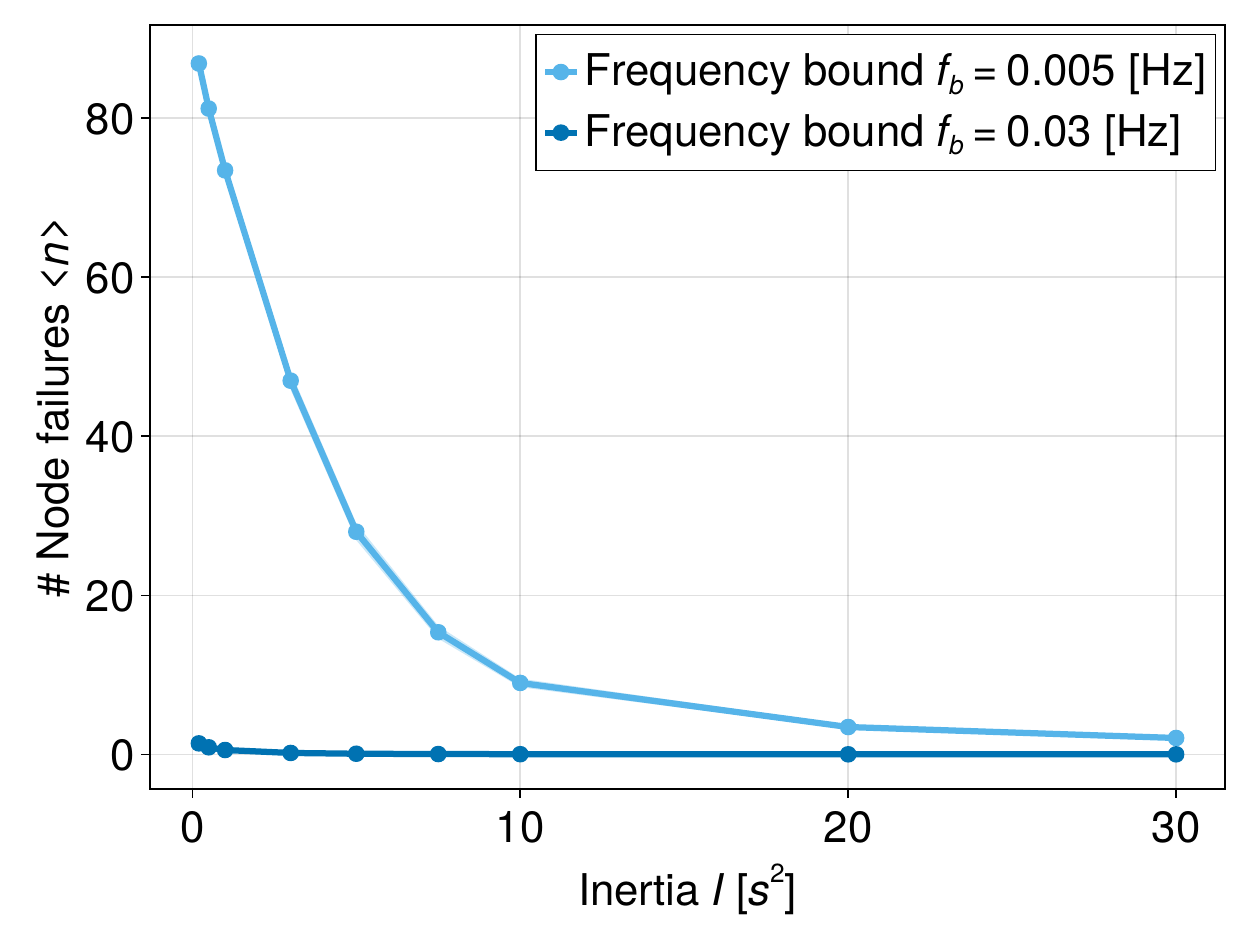}
  \caption{
  Number of node failures $\langle n\rangle$ as a function of inertia $I$ for two different frequency bounds $f_b$. The Watts-Strogatz networks are analyzed, where the full-failure model allows for both node and line failures. The standard error is shown in shaded bands. The number of node failures $\langle n\rangle$ decreases with increasing frequency bounds $f_b$.}
  \label{fig:WS_nodes_only_different_f_b}
\end{figure}

The same qualitative behavior appears when the fractions of node failures $\frac{\langle n\rangle}{N}$ and line failures $\frac{\langle l\rangle}{M}$ are considered separately in Figs.~\ref{fig:f_b_vs_failures}c and~\ref{fig:f_b_vs_failures}d. However, if we study the node-failure-only model, the number of node failures decreases (Supplementary Fig.~\ref{fig:WS_nodes_only_different_f_b}), as naively expected. This mirrors results in the node-failure-only analysis of~\cite{parkOptimalLocationReinforced2025} and indicates that the increase in failures $\langle F\rangle$ originates from the coupling of node and line failures, rather than from either individually.
In Fig.~\ref{fig:f_b_vs_failures}, inertia is set to $I = \SI{30.0}{\second\squared}$ for the WS networks and the inertia scaling for the power grid model is set to $8.0$. The effect persists, but is less pronounced for lower inertia values, see Supplementary Figs.~\ref{fig:WS_f_b_vs_failures} and~\ref{fig:RTS_f_b_vs_failures}.

\begin{figure*}
\includegraphics[width=0.95\textwidth,height=\textheight,keepaspectratio]{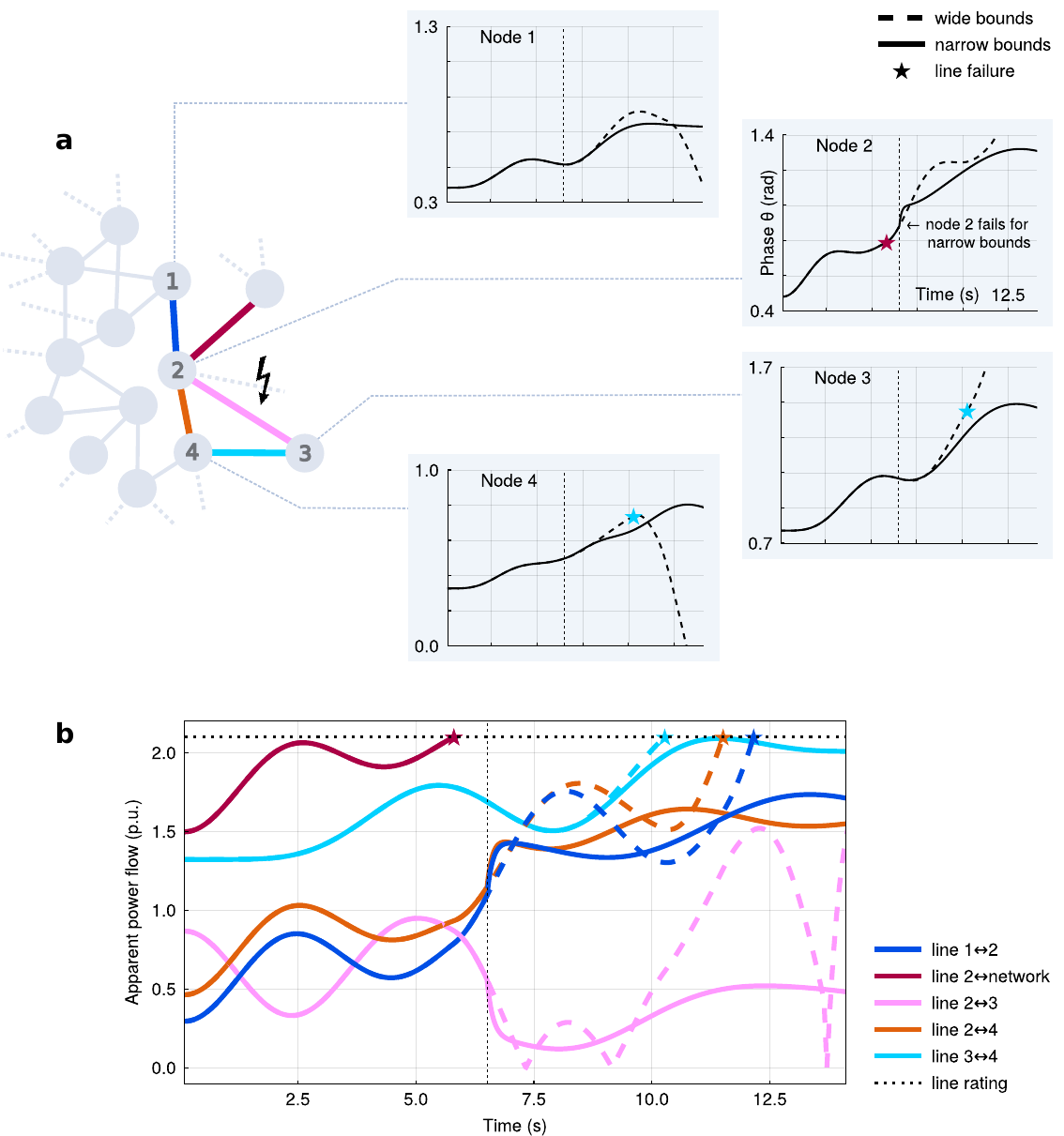}
\caption{%
An illustrative example for the dynamical Braess's paradox. Early node failure, induced by narrow frequency bounds, significantly reduces line failures through fast local adaptation of power flows.
\textbf{a} Cut-out of a 100 node Watts–Strogatz network (dashed gray lines indicate connection to network). To trigger a cascade, the line marked by a lightning bolt is removed at $t = \SI{0.1}{\second}$.  We compare the transient response for wide (dashed) and narrow (solid) frequency bounds. Panels show the phase angle $\theta$ of the indicated nodes. The color-coded lines in the network correspond to the power flows shown in \textbf{b} that are directly determined by the phase angle differences.
For narrow bounds, node 2 exceeds the frequency bounds at $t = \SI{6.51}{\second}$ and fails. As a consequence, the inertia at this node is removed and the local power surplus is slightly reduced. 
Due to the lack of inertia and power surplus, the phase angle of node 2 in \textbf{a} (solid line) reacts much faster to the local power imbalance and shows only a small phase overshoot compared to the inert phase response (dashed line). Compared to the wide-bound scenario, this leads to significantly reduced power-flow overshoots and power flows on the adjacent lines (1,2) and (2,4) and line (3,4) in \textbf{b} remain within their capacity limits (line rating). This prevents further line overloads and suppresses a larger cascade. In contrast, wide bounds result in pronounced overshoots and the onset of a large cascading failure involving multiple additional line and node outages. Only the initial stage of this cascade is shown.}
\label{fig:WS_illustrative_fig}  
\end{figure*}

In Fig.~\ref{fig:WS_illustrative_fig}, we present an illustrative example of this effect for a single cascading event. To trigger a cascade, the line marked by a lightning bolt is removed at $t = \SI{0.1}{\second}$, which causes the failure of line (2,network). This line failure causes a power imbalance at node~2. For narrow bounds, this causes node~2 to exceed its frequency limit at $t=\SI{6.51}{\second}$ and fail, removing its inertia and the grid-forming part of its power injection. The phase angle of node~2 in Fig.~\ref{fig:WS_illustrative_fig}a (solid line) reacts much faster to the local power imbalance and shows only a small phase overshoot compared to the inertial phase response (dashed line). The flows are determined by the phase angle differences between the respective nodes, according to Eqs. (\ref{eq:active_power_flow_single_line}) and (\ref{eq:reactive_power_flow_single_line}). Consequently, the power flows on the adjacent lines (1,2),(2,4) and (3,4) in Fig.~\ref{fig:WS_illustrative_fig}b remain within their capacity limits. In contrast, wide bounds result in pronounced overshoots that lead to the failure of lines (3,4), (1,2) and (2,4), and the onset of a large cascading failure involving multiple additional line and node outages. Hence, we observe a clear dynamical Braess's paradox: A more robust node leads to increased vulnerability.

\subsection{Mechanisms underlying the dynamical Braess's paradox} \label{sec:mechanisms_braess}

The concrete example in Fig.~\ref{fig:WS_illustrative_fig} suggests a mechanism for the dynamical Braess's paradox: Locally reduced inertia leads to a fast equilibration of the local phase dynamics, preventing overshoots. A second plausible effect is that lines connected to a node injecting power will more frequently transport power away from a node. The failure of these lines will lead to a local power surplus, and the failure of power injection at the node will reduce or remove this surplus.

\begin{figure} \includegraphics[width=\linewidth]{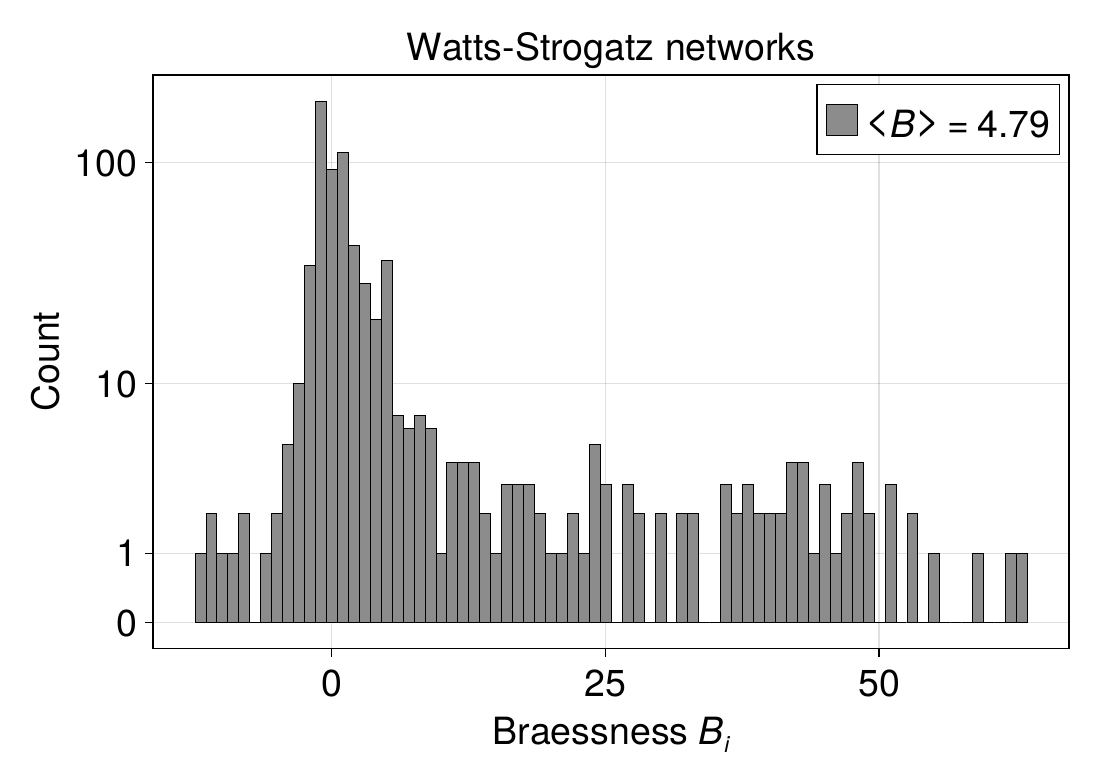}
  \caption{Histogram of Braessness for the Watts–Strogatz ensemble, with an average Braessness of $\langle B\rangle =4.79$. A significant fraction of individual cascades shows a high Braessness. For $52\%$ of the cascades, $B>0$; for $35\%$, $B<0$; and for the remaining $13\%$, $B=0$.
  }
  \label{fig:WS_braessness_histogramm_lines+nodes}
\end{figure}

To investigate these two mechanisms, we will analyze the individual cascades. Recall that $F_i$ denotes the number of failures in the cascade triggered by initial failure at~$i$. We define the strength of Braess's paradox for the cascade triggered by $i$ as Braessness $B_i = F^{\text{wide}}_i - F^{\text{narrow}}_i$, with $f_b^{\text{wide}} > f_b^{\text{narrow}}$, while keeping all other parameters fixed. We only consider cascades that include secondary failures, $F_i > 1$.
In Fig.~\ref{fig:WS_braessness_histogramm_lines+nodes}, we provide the distribution of Braessness
in the WS ensemble. A considerable number of cascades feature high $B_i$, with differences in cascade size for narrow and wide frequency bounds above 50. For $52\%$ of the cascades it is $B>0$. Conversely, only $35\%$ of the cascades exhibit the naively expected behavior, where more robust nodes lead to smaller cascades ($B<0$), while for $13\%$ of the cascades it is $B=0$.
The Braessness for the power grid that shows a similar behavior and $B$ for node and line failures separately are shown in Supplementary Fig.~\ref{fig:histograms_braessness_line_node_failures_separately}. We find that high $B$ values are caused by high Braessness of line failures.

To study the two plausible mechanisms behind the Braess behavior, we introduce two variants of our node failure model (full-failure model) that aim to isolate both effects.

i) The first mechanism concerns faster equilibration.
When a line %
that is connected to a node fails, a local power imbalance at this node results. The power imbalance at the node then leads to an increase or decrease in frequency. 
Without inertia, the phase angle of the node reacts much faster to the local power imbalance and shows only a small overshoot compared to the inertial phase response and the large overshoot that can occur with inertia. To isolate this effect we consider the inertia-failure model, where node failure means that the inertia is set to zero, while the power injection remains unchanged. We expect the inertia-failure model to capture the faster equilibration and reduction of overshoot we observed in Fig.~\ref{fig:WS_illustrative_fig}.

ii) The second mechanism concerns reduced power imbalance.
The set of nodal power injections across the network determines the steady-state power flows on the lines. We expect the lines at a node injecting power to transport power away from that node, establishing a correlation between a node’s power injection and the loading of its incident lines.
If a node injects power, flows on connected lines are on average directed away from it. Conversely, if the node consumes power, they are directed towards it on average. The failure of a line creates a local power imbalance at the adjacent node. When a node fails, the grid-forming injection is set to zero, which, on average, reduces this imbalance. To isolate this effect we introduce the power-failure model, where node failures only modify the power injection but keep inertia unchanged. We expect this model to capture the impact of reduced power imbalance, and thus reduced frequency deviations and line loadings.

Using these models, we evaluated the individual effect of both the inertia- and the power-failure mechanism. The results show that both effects play a decisive role in the emergence of Braess's paradox.
We compared the individual line Braessness of the full-failure model $B_i$, the inertia-failure model $B^{\text{Inertia}}_i$ and the power-failure model $B^{\text{Power}}_i$. 
\begin{figure} %
  \includegraphics[width=\linewidth]{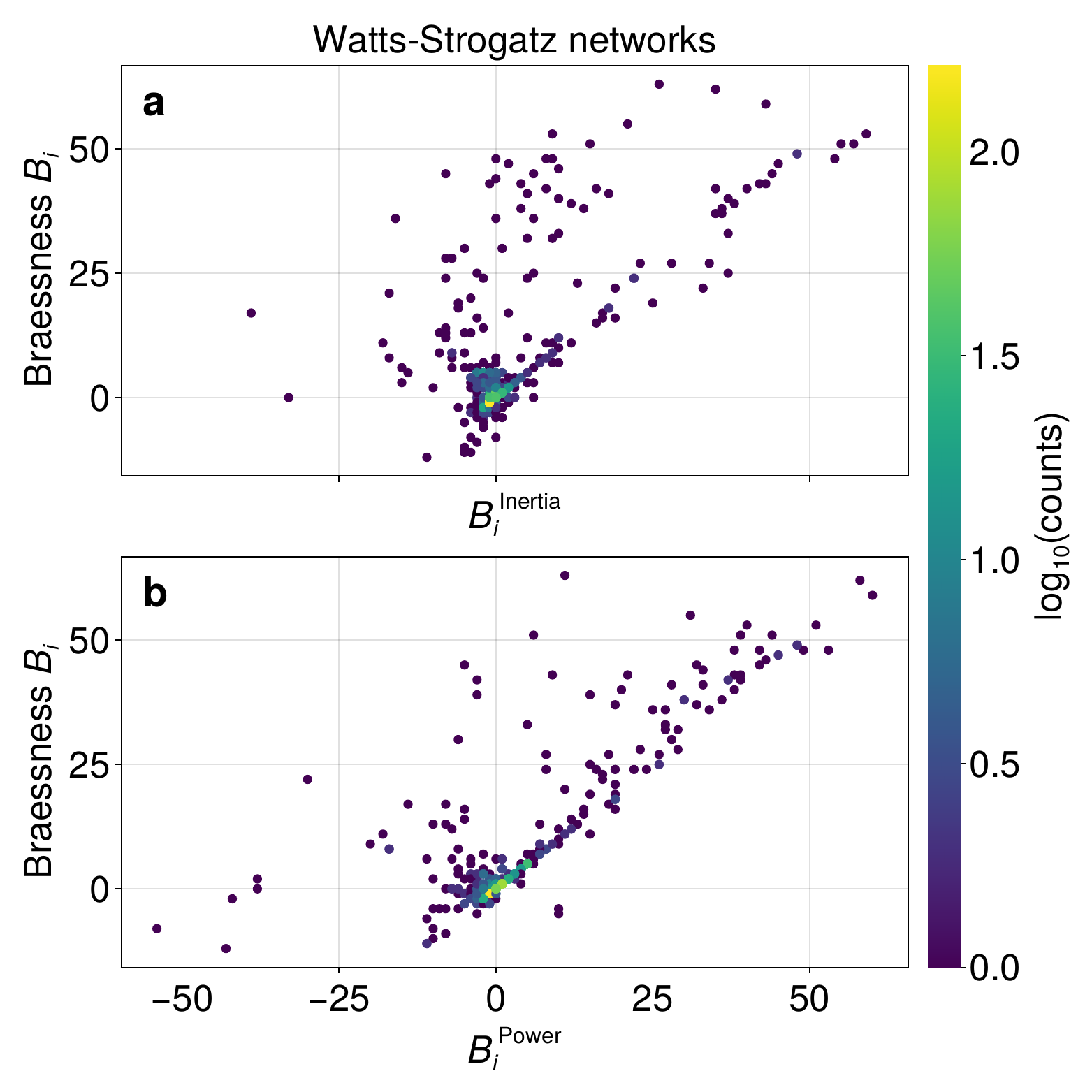}
  \caption{Braessness $B_i$ of individual cascades in the Watts–Strogatz ensemble for the full-failure model, as well as for the two mechanisms occurring during node failure: inertia-failure $B^{\text{Inertia}}_i$ and power-failure $B^{\text{Power}}_i$. \textbf{a} and \textbf{b} show that a significant number of individual cascades show a high Braessness for both inertia-failure and power-failure mechanisms separately. For separate mechanisms, highly ``anti-Braess'' cascades, which are absent in the full-failure model, are also present. Cascades that exhibit positive Braessness for either mechanism do not experience a reduction of Braessness from the other mechanism. The full Braess's paradox we observe arises from a non-linear interaction of both mechanisms.  
   }
  \label{fig:WS_braessness_vs_braessness_2D_count_lines+nodes}
\end{figure}
Figs.~\ref{fig:WS_braessness_vs_braessness_2D_count_lines+nodes}a and~\ref{fig:WS_braessness_vs_braessness_2D_count_lines+nodes}b show that a significant amount of individual cascades has a high Braessness for the inertia-failure and the power-failure mechanism, separately. This emphasizes that both effects play an important role in Braess's paradox.
However, for both individual mechanisms we also observe a number of individual cascades with high anti-Braessness, that is $B < 0$.

When both mechanisms act together in the full-failure model, their interaction is highly non-linear. Despite both mechanisms exhibiting considerable and strong anti-Braessness, cascades in the full-failure model almost always have higher Braessness than in each individual model. Strongly anti-Braess cascades ($B < -20$), present for each mechanism in isolation, completely disappear in the full-failure model.
As a result, Braess's behavior does not just occur, but dominates, while anti-Braessness disappears almost completely in these grid ensembles.

\begin{figure} 
  \includegraphics[width=0.5\linewidth]{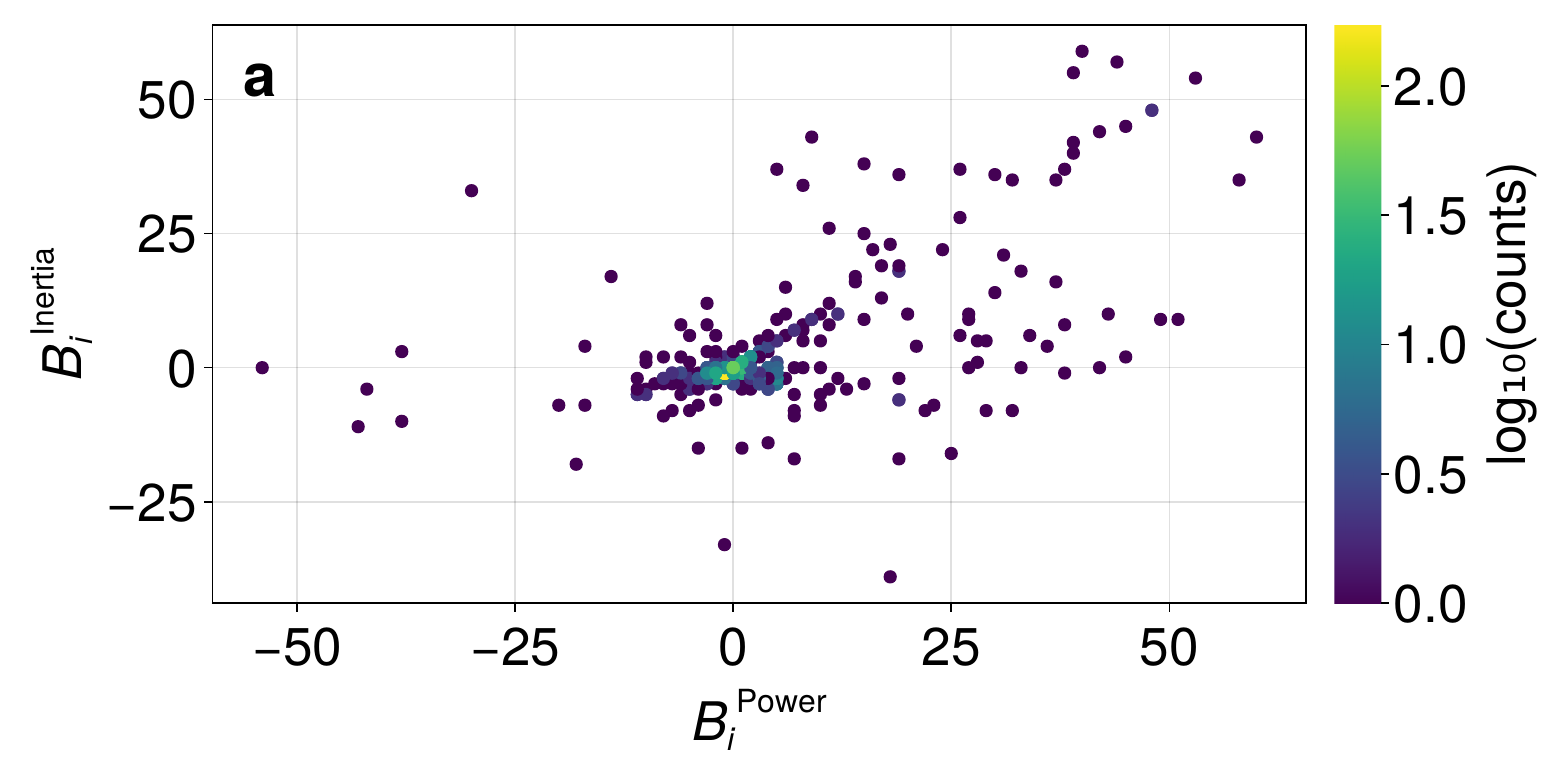}
  \includegraphics[width=0.5\linewidth]{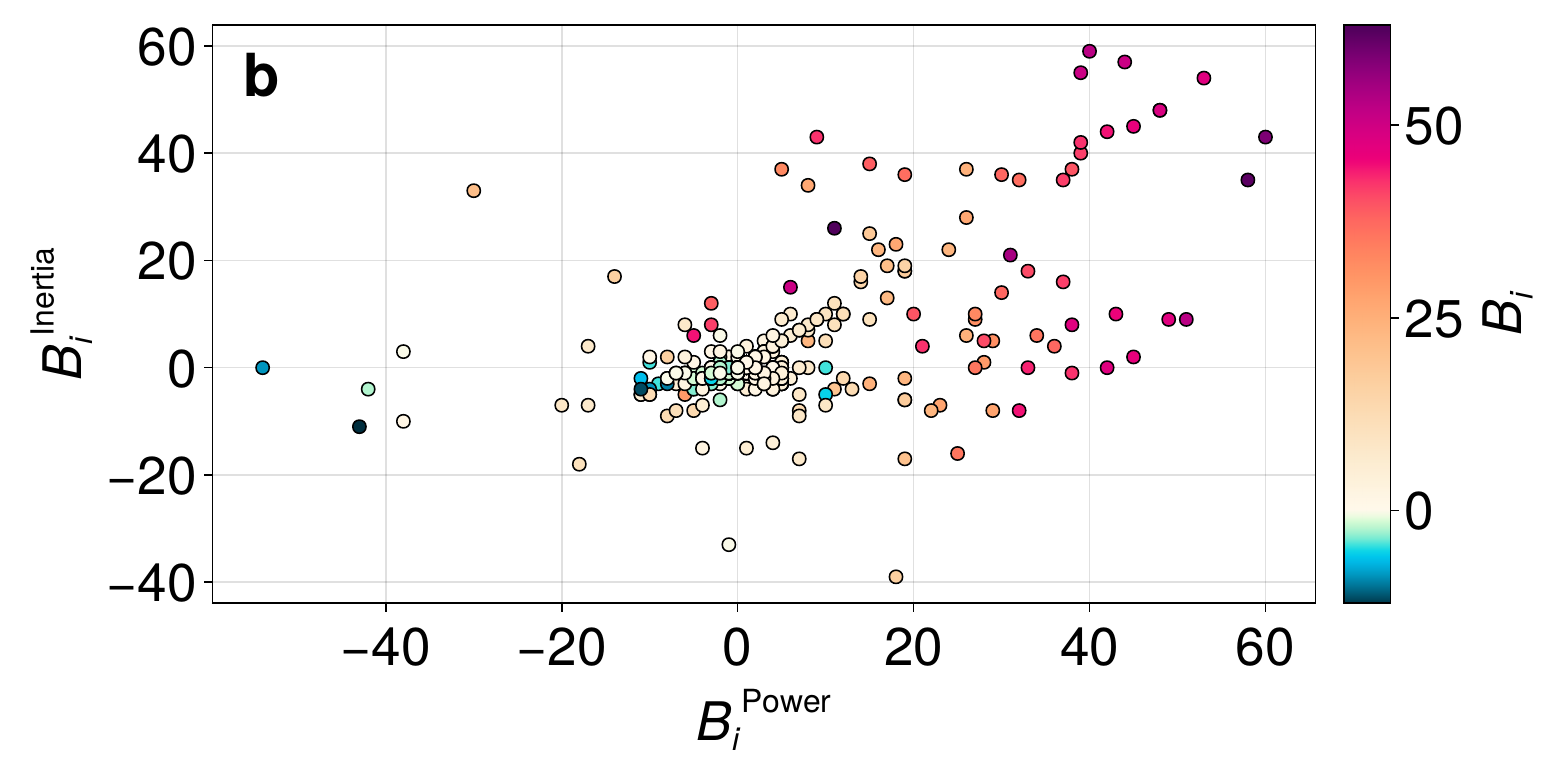}
  \caption{Braessness $B_i$ of individual cascades in the Watts–Strogatz ensemble for the same data as in Fig.~\ref{fig:WS_braessness_vs_braessness_2D_count_lines+nodes} for different representations.
  \textbf{a} shows the relation between $B^{\text{Power}}$ and $B^{\text{Inertia}}$.
  \textbf{b} shows all three mechanisms (full-failure, inertia-failure and power-failure model) with the full-failure model color-coded. We observe that for opposing signs of $B^{\text{Power}}$ and $B^{\text{Inertia}}$ (in the second and fourth quadrants) Braess's behavior dominates, i.e. $B_i>0$. Consequently, both the power-failure and the inertia-failure mechanisms can separately lead to Braess's paradox.}
  \label{fig:WS_braessness_power_vs_inertia_vs_full_2D_lines+nodes}
\end{figure}

Supplementary Fig.~\ref{fig:WS_braessness_power_vs_inertia_vs_full_2D_lines+nodes}a shows the relation between $B^{\text{Power}}$ and $B^{\text{Inertia}}$, and Supplementary Fig.~\ref{fig:WS_braessness_power_vs_inertia_vs_full_2D_lines+nodes}b provides the relation between $B^{\text{Power}}$ and $B^{\text{Inertia}}$ with $B_i$ color-coded. The Braessness for node and line failures separately is given in Supplementary Fig.\ref{fig:WS_braessness_vs_braessness_2D_count_lines_and_nodes_separately}.

\begin{figure}
  \includegraphics[width=0.5\linewidth]{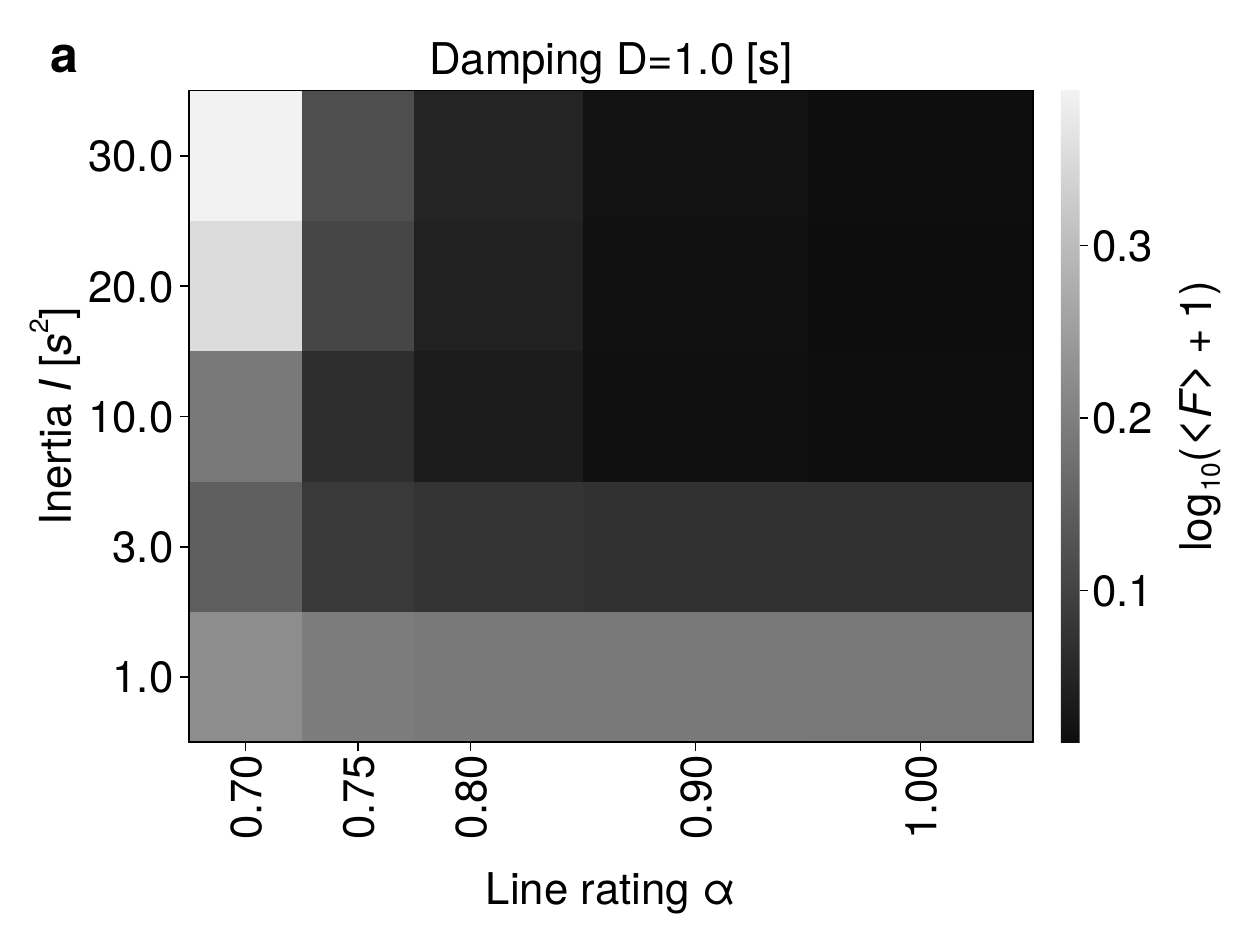}
  \includegraphics[width=0.5\linewidth]{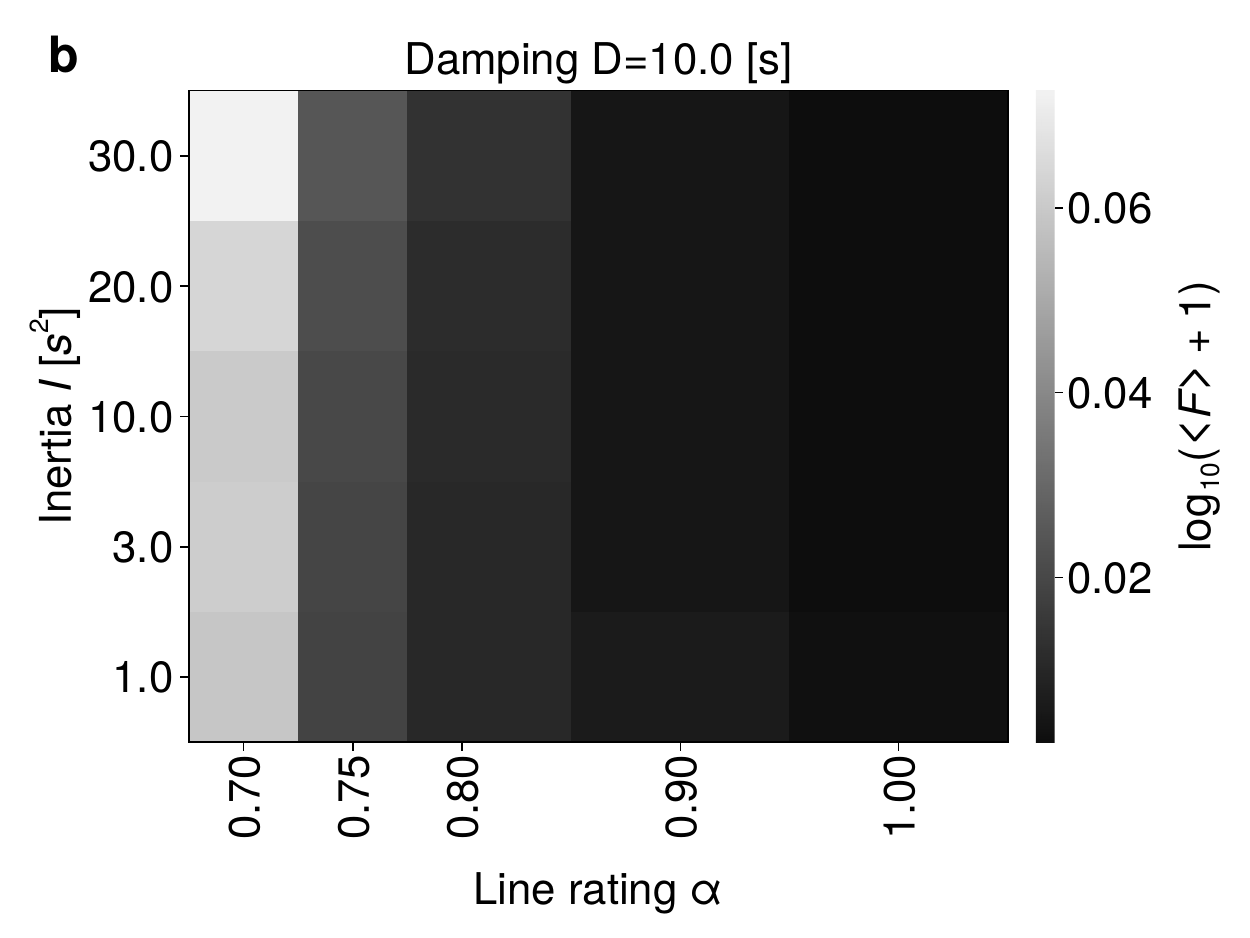}
  \caption{Number of failures $\langle F\rangle$ for the Watts-Strogatz networks for different values of the safety parameter $\alpha$ and different inertia, where the full-failure model allows for both node and line failures with a frequency bound $f_b=0.03$. In \textbf{a} $D=\SI{1}{\second}$ and in \textbf{b}, $D=\SI{10}{\second}$. Note that the y-axes are nonlinearly scaled. Increasing $\alpha$, and thus enhancing line robustness, does not reduce overall grid resilience in \textbf{a} and \textbf{b}. Increasing $\alpha$ reduces the number of failures.}
  \label{fig:alpha_variation}
\end{figure}

Finally, we investigated whether similar effects occur when the safety parameter $\alpha$ introduced in Eq. (\ref{eq:line_failure_condition}) varies. Specifically, we tested whether increasing $\alpha$, and thus enhancing line robustness, could reduce overall grid resilience. Supplementary Fig.~\ref{fig:alpha_variation} shows that this is not the case: increasing $\alpha$ reduces the number of failures. Schäfer et al.~\cite{schaferDynamicallyInducedCascading2018}, who studied dynamic line failures without node failures, found a similar result.

\section{Discussion} %
Cascading failures in real-world power grids are driven by a complex interplay of node and line failures. We coupled node and line failures in a dynamic cascading failure model, and found that this interplay leads to novel phenomena.

The analysis of varying inertia and frequency bounds shows that neither higher inertia nor wider frequency bounds necessarily improve grid stability. Instead, both can paradoxically increase the power grid's vulnerability to cascading failures. We discovered a dynamical Braess's paradox, where increasing the nodal robustness can decrease the grid's resilience to cascading failures. We further identified the mechanisms underlying the observed novel phenomena for both cases. The rapid local equilibration of the flow due to inertia failure is of particular importance here. Inertia of the power system is classically considered as a global, aggregated quantity, governing the overall rate of change of the frequency (RoCoF) in the grid. Its impact on the local equilibration dynamics has not been widely considered so far.

While the specific parameter values used in this study are not intended to represent any particular real-world power grid, the qualitative insights are of high practical relevance. In particular, the findings regarding inertia may inform future control strategies, as the virtual inertia of inverters interfacing renewable energy sources with the grid can be tuned. We saw that a naive increase in inertia, without carefully considering further dynamic parameters, can be problematic. The dynamical Braess's paradox is particularly relevant in highly decentralized energy systems, in which inertia providing actors are no longer connected at the highest grid level, but in distribution grids with comparatively poorly understood dynamics. Local destabilizing dynamic phenomena such as the mechanisms underlying our observed Braess's paradox must be avoided in a principled way.
Technical constraints generally prevent the widening of generator frequency bounds $f_b$, but less so for power electronic devices. Narrowing them to improve system stability may be feasible and could represent a practical measure to enhance grid resilience.

To arrive at actionable recommendations for real power grids, a more realistic underlying dynamical model that captures the characteristics of future power grid actors is required~\cite{koglerNormalFormGridForming2022, buttnerComplexPhaseDataDrivenIdentification2025}. Voltage, current and RoCoF protections would need to be considered~\cite{hossainRobustControlGrid2014, fanNetworkbasedStructurepreservingDynamical2022}.
Voltage dynamics that are not present in the Bergen-Hill model play a central role in many observed grid outages, including the Iberian Peninsula blackout of 2025~\cite{entso-eFinalReportGrid2026}. However, some case studies of cascading failures in intermediate complexity models have also observed that adding protection devices that deliberately fail components can reduce cascade size~\cite{fanNetworkbasedStructurepreservingDynamical2022}. This suggests that similar mechanisms to those we discovered play a role in detailed engineering style models, too.

From a purely theoretical perspective, the interplay of nodal and line dynamics has only recently been fully explored, with many notable recent results~\cite{bernerAdaptiveDynamicalNetworks2023}. Our work initiates the study of cascades in such systems. While our analysis was motivated by phenomena in electrical power grids, the underlying mechanisms are applicable to a broader class of flow networks. The essential characteristics are (i) node- and line-level thresholds that trigger operational shutdown or protection events, and (ii) fast transient flow redistributions following local disturbances. When these features coexist, the coupling between node and line failures can give rise to collective cascade dynamics similar to those observed here. Consequently, our findings may qualitatively extend to other supply networks, such as gas networks, where transient dynamics and protection limits interact. Beyond supply networks, our results also suggest that node and line cascades in more general adaptive networks will have substantially different characteristics than their line/node-failure-only counterparts, which should be explored in depth.

\section{Methods}

\subsection{Power flow on lines} \label{sec:apparent_power}
We use the complex representation of voltage $v$ and current $i$. The current between nodes $k$ and $m$ in an AC circuit follows Ohm's law:
\begin{equation} \label{eq:ohms_law}
  i_{km} = y_{km} (v_k - v_m),
\end{equation}
where $y_{km} = z_{km}^{-1}$ is the admittance and $z_{km} = r_{km} + j x_{km}$ the impedance with resistance $r_{km}$, reactance $x_{km}$ and the imaginary unit $j$. The admittance matrix $Y$ encodes the network topology: For $y_{km}=0$ there is no line between nodes $k$ and $m$.
The apparent power flow is defined as:
\begin{equation} \label{eq:apparent_power_flow}
  S_{km} = v_k i_{km}^* = P_{km} + j Q_{km},
\end{equation}
where $P_{km}$ and $Q_{km}$ are the active and reactive power flow.
The resistance in high voltage transmission lines is negligible, thus we assume $r_{km} \approx 0$. With the voltage magnitude $V_k$ and phase angle $\theta_k$ of $v_k$, we get the active and reactive power flow on a line:
\begin{align}
\label{eq:active_power_flow_single_line}
  & P_{km}= \frac{\left|V_k\right|\left|V_m\right|}{x_{km}} \sin (\theta_k - \theta_m), \\
\label{eq:reactive_power_flow_single_line}
  & Q_{km}= \frac{1}{x_{km}} \left( \left|V_k\right|\left|V_m\right|  \cos (\theta_k - \theta_m) - \left|V_k\right|^2 \right).
\end{align}
We finally define the absolute apparent power flow as:
\begin{equation}
  \label{eq:maximum_apparent_power}
  |S_{km}|=|P_{km} + jQ_{km}|,
\end{equation}
In this work, we further assume the voltage magnitude to be time-independent of $V_k=1$. This assumption is valid as long as we model short time-scales. With this assumption, we obtain:
\begin{equation}
  \label{eq:apparent_power_sin}
  |i_{km}| = |S_{km}|= \frac{2}{x_{km}}\left|\sin\left(\frac{\theta_k - \theta_m}{2}\right)\right|.
\end{equation}

\subsection{Balancing of power injections} \label{sec:methods_balancing}
To obtain an initially stable state of the system, where all machines run in synchrony at the reference frequency, we need a balanced system, that is $\sum_{k=1}^{N}P_k = 0$.  When starting from unbalanced power injections $\tilde{P}_k$, we balance them by applying the following to each grid-forming node~$k$:
\begin{equation}
  \label{eq:balancing}
  P_k^{\text{GFM}} = \tilde{P}_k^{\text{GFM}} - \frac{|\tilde{P}_k^{\text{GFM}}|}{\sum_{m=1}^{N}|\tilde{P}_m^{\text{GFM}}|} \sum_{m=1}^{N}\tilde{P}_m.
\end{equation}

\subsection{Watts–Strogatz networks: Choice of rewiring probability \texorpdfstring{$\beta$}{beta}}\label{sec:methods_WS_choice}
In the WS graph generation model, each node is initially connected to its four nearest neighbors (mean degree of $k = 4$) on a ring, after which each line is independently rewired with probability $\beta$. Varying $\beta$ interpolates between a regular ring lattice ($\beta = 0$) and a fully randomized network ($\beta = 1$), thereby generating different network topologies (see Supplementary Fig.~\ref{fig:WS_example_topologies}). In Supplementary Fig.~\ref{fig:WS_lines_only_different_beta}, each curve corresponds to a different rewiring probability $\beta$, with larger values of $\beta$ leading to fewer line failures. For $\beta=0$, the WS network forms a ring lattice with a one-dimensional structure. A single line failure in this structure reduces the flow capacity by one third, potentially leading to a bottleneck that increases the cascading potential. As $\beta$ increases, shortcuts are introduced that reduce potential structural bottlenecks, leading to fewer cascades. We observed the same qualitative behavior for different values of $\beta$ in Fig.~\ref{fig:inertia_effect}a. As a result of increasing $\beta$, the minimum in the number of failures at intermediate frequency bounds shifts to higher inertia values. For $\beta > 0.5$, no further reduction in cascadability is observed. We therefore use $\beta = 0.5$ as a representative value for a generic network topology.

\subsection{Parameter choice: Frequency bounds \texorpdfstring{$f_b$}{fb} and inertia \texorpdfstring{$I$}{I}} \label{sec:methods_parameter_choice}

\begin{figure}
  \includegraphics[width=0.5\linewidth]{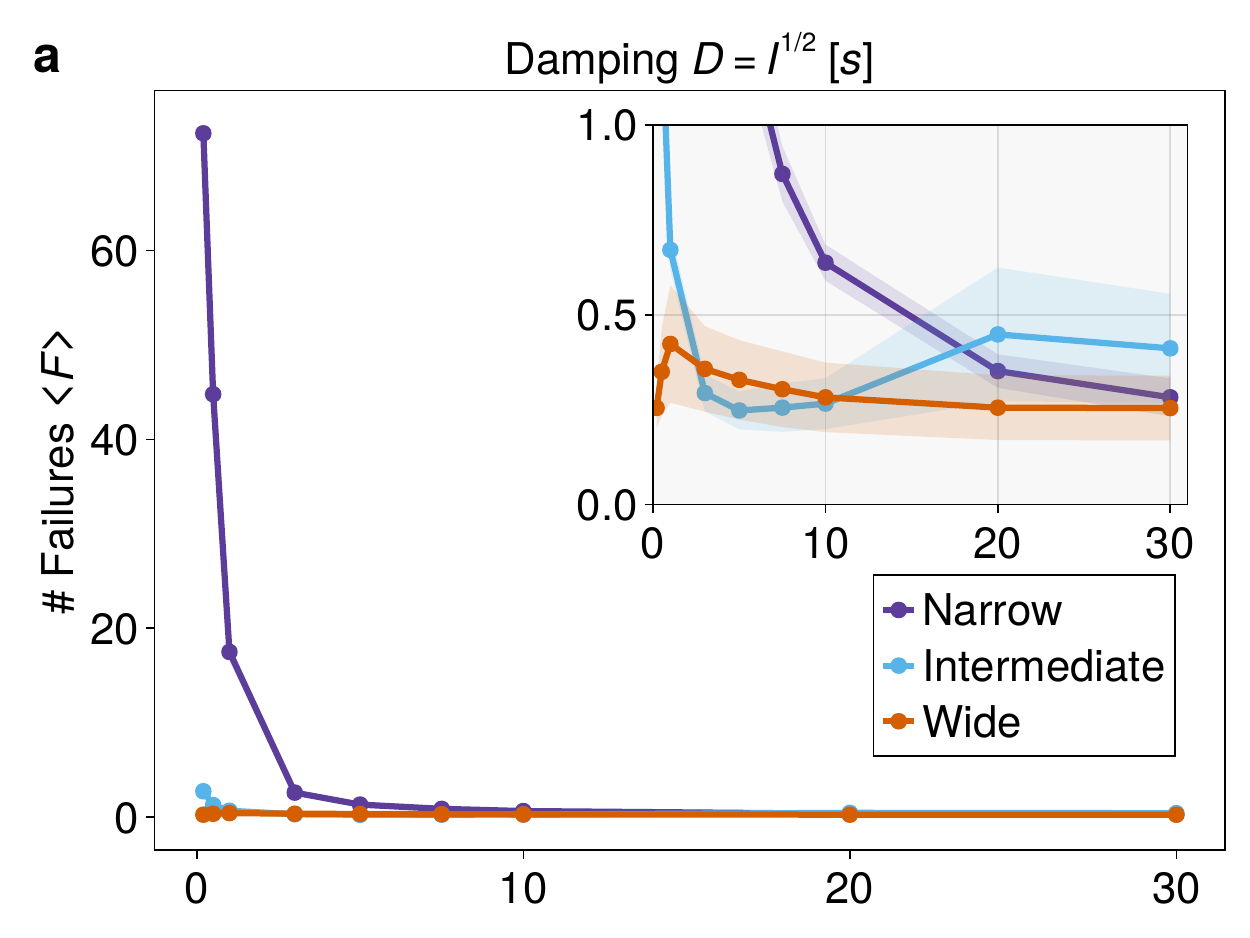}
  \includegraphics[width=0.5\linewidth]{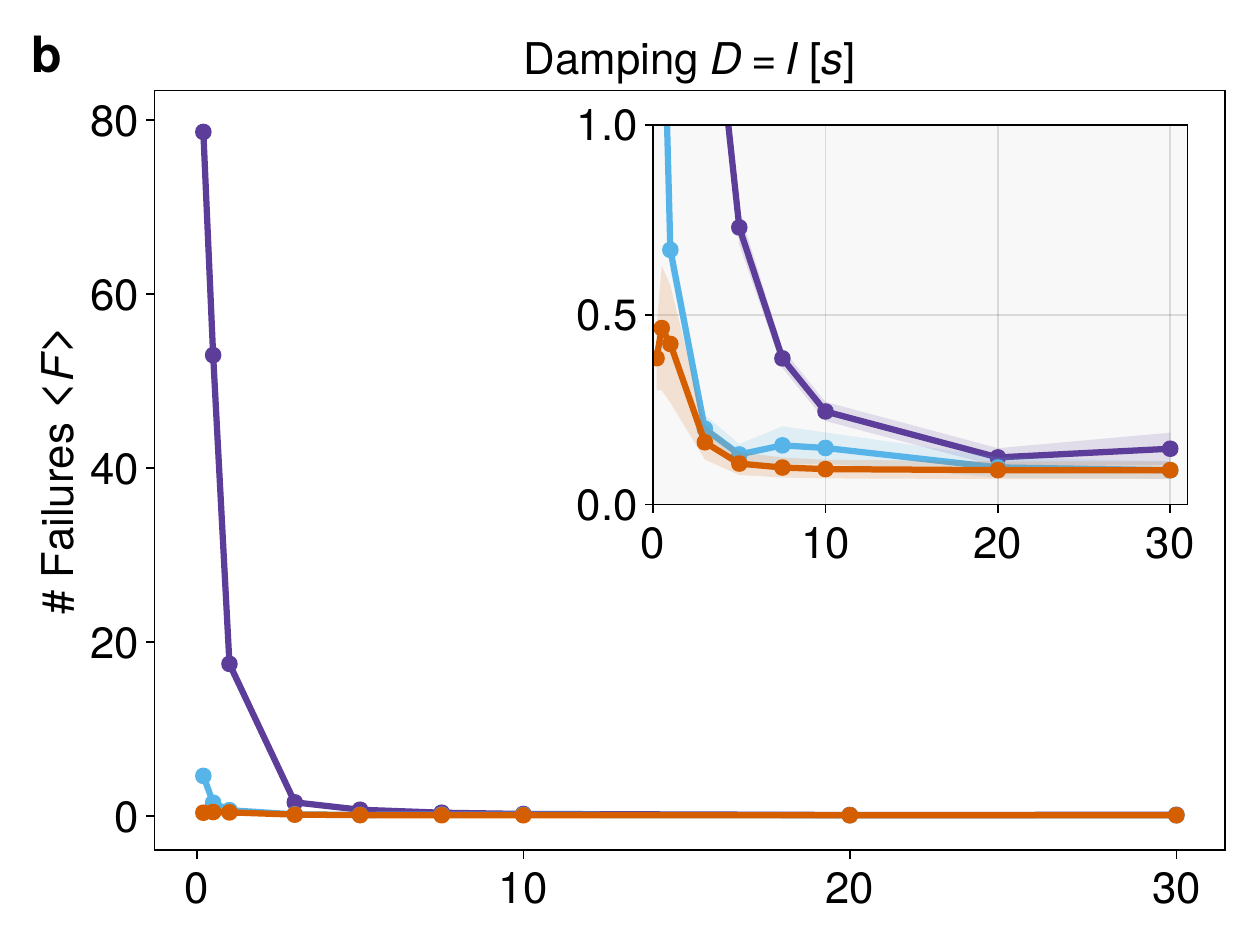}
  \\
  \includegraphics[width=0.5\linewidth]{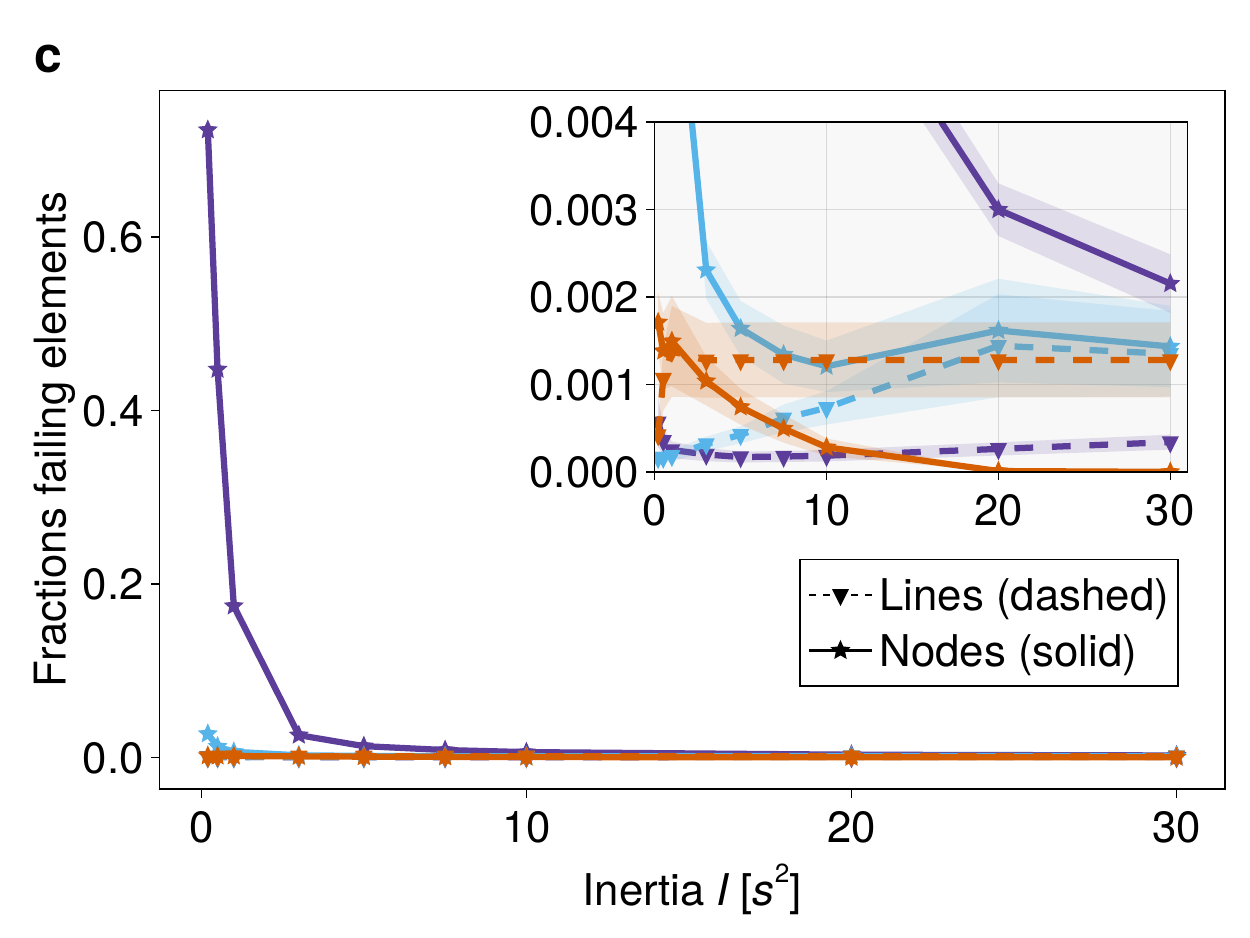}
  \includegraphics[width=0.5\linewidth]{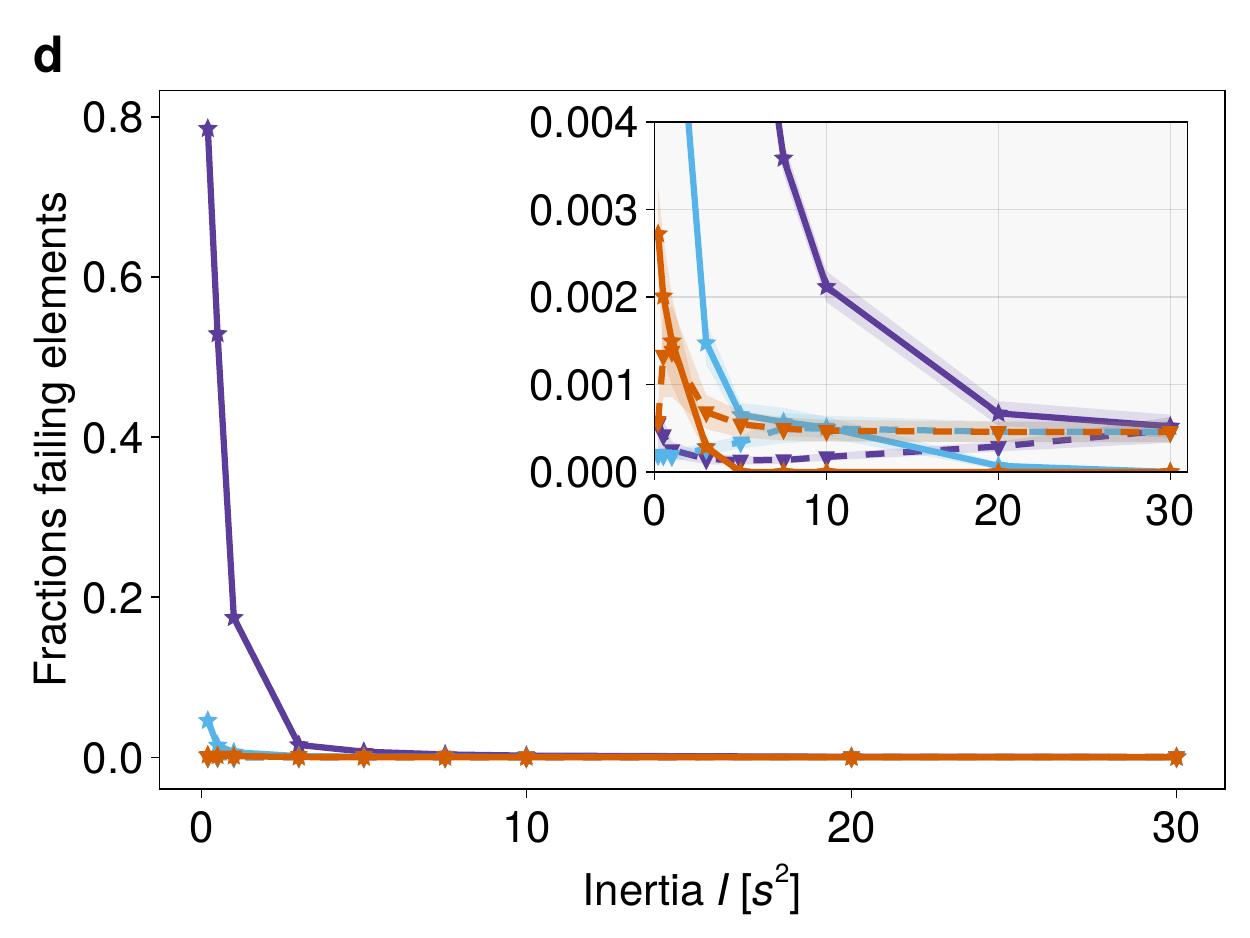} 
  \caption{Failures as a function of inertia $I$ with damping $D=I^{1/2}$ in \textbf{a,c} and $D=I$ in \textbf{b,d} for different frequency bounds. Watts-Strogatz networks are analyzed, and the full-failure model allows for both node and line failures. The standard error is shown in shaded bands.
  \textbf{a,b} show the number of failures $\langle F\rangle$, while \textbf{c,d} show the fractions of node failures $\frac{\langle n\rangle}{N}$ (solid) and line failures $\frac{\langle l\rangle}{M}$ (dashed) separately.
  For $D = I^{1/2}$ in \textbf{a}, $\langle F \rangle$ is lower than for $D = \SI{1}{\second}$ in Fig.~\ref{fig:inertia_effect}a. For $D = I$ in \textbf{b}, $\langle F \rangle$ decreases further.
  Moreover, we observe that compared to setting $D = \SI{1}{\second}$, in \textbf{a,b} the qualitative behavior for narrow bounds changes: $\langle F \rangle$ tends to decrease with inertia. Thus, scaling inertia and damping jointly can eliminate the problem of line-induced overloads.}
  \label{fig:WS_uebergang_lines+nodes_different_scalings}
\end{figure}

\begin{figure}
  \includegraphics[width=0.33\linewidth]{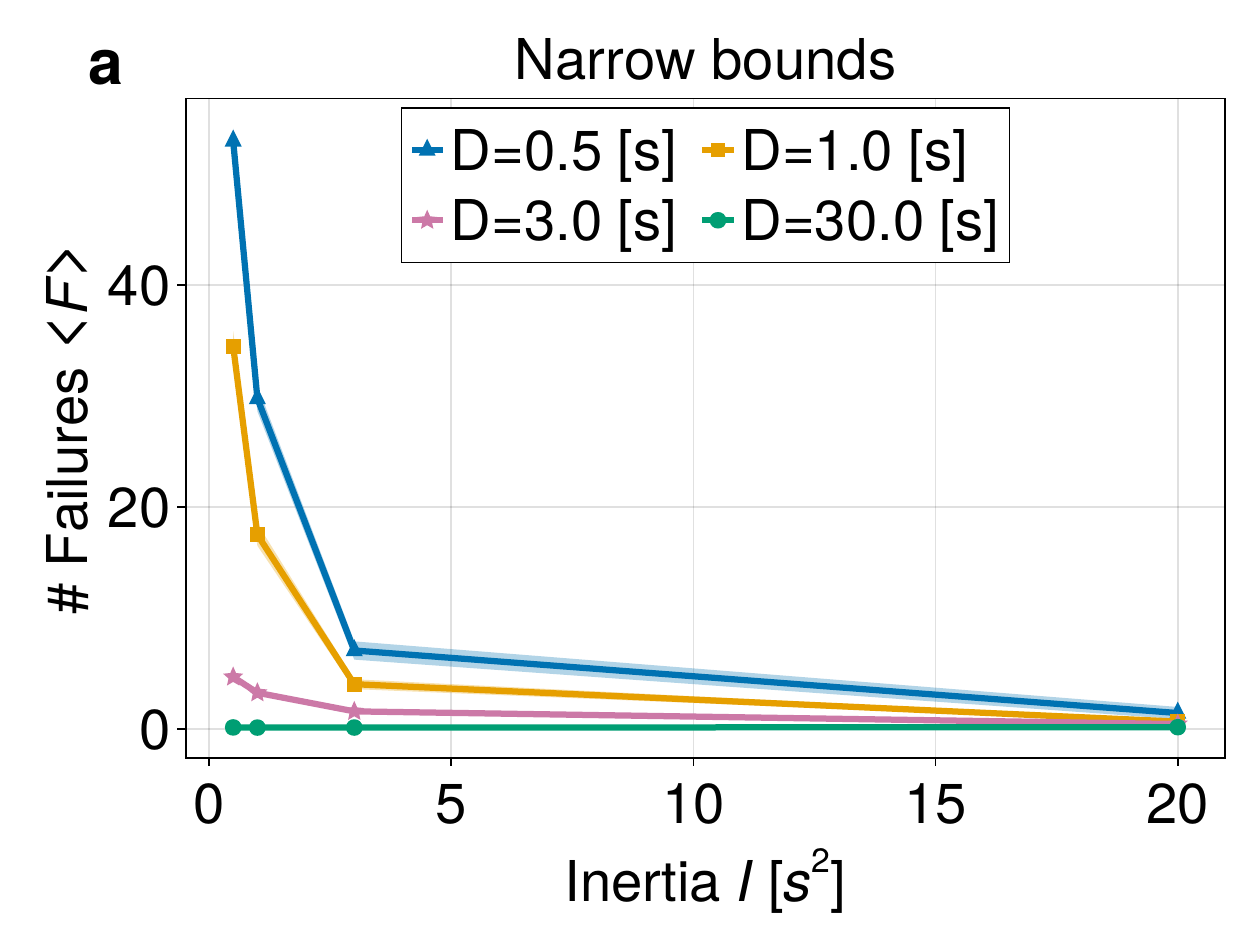}
  \includegraphics[width=0.33\linewidth]{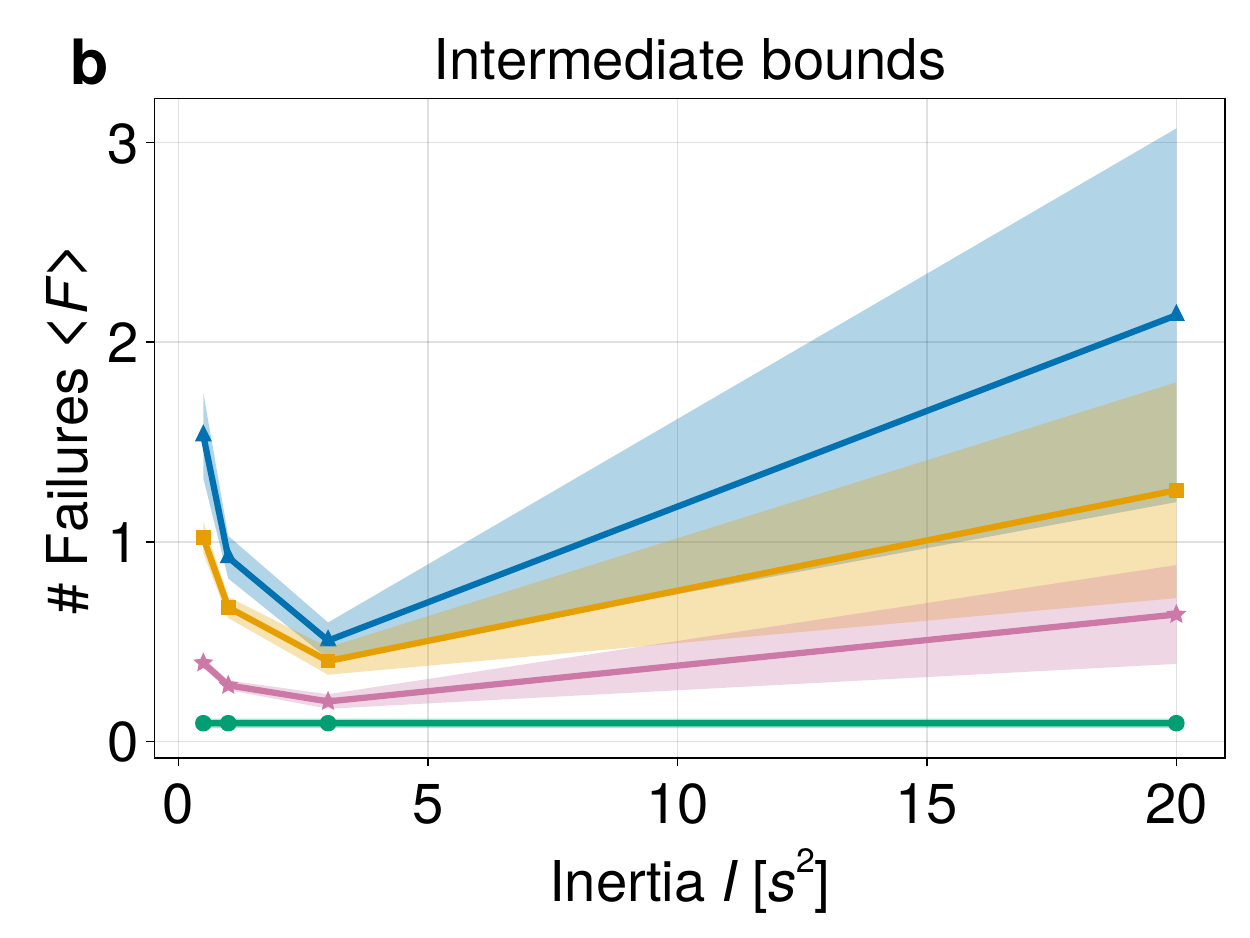}
  \includegraphics[width=0.33\linewidth]{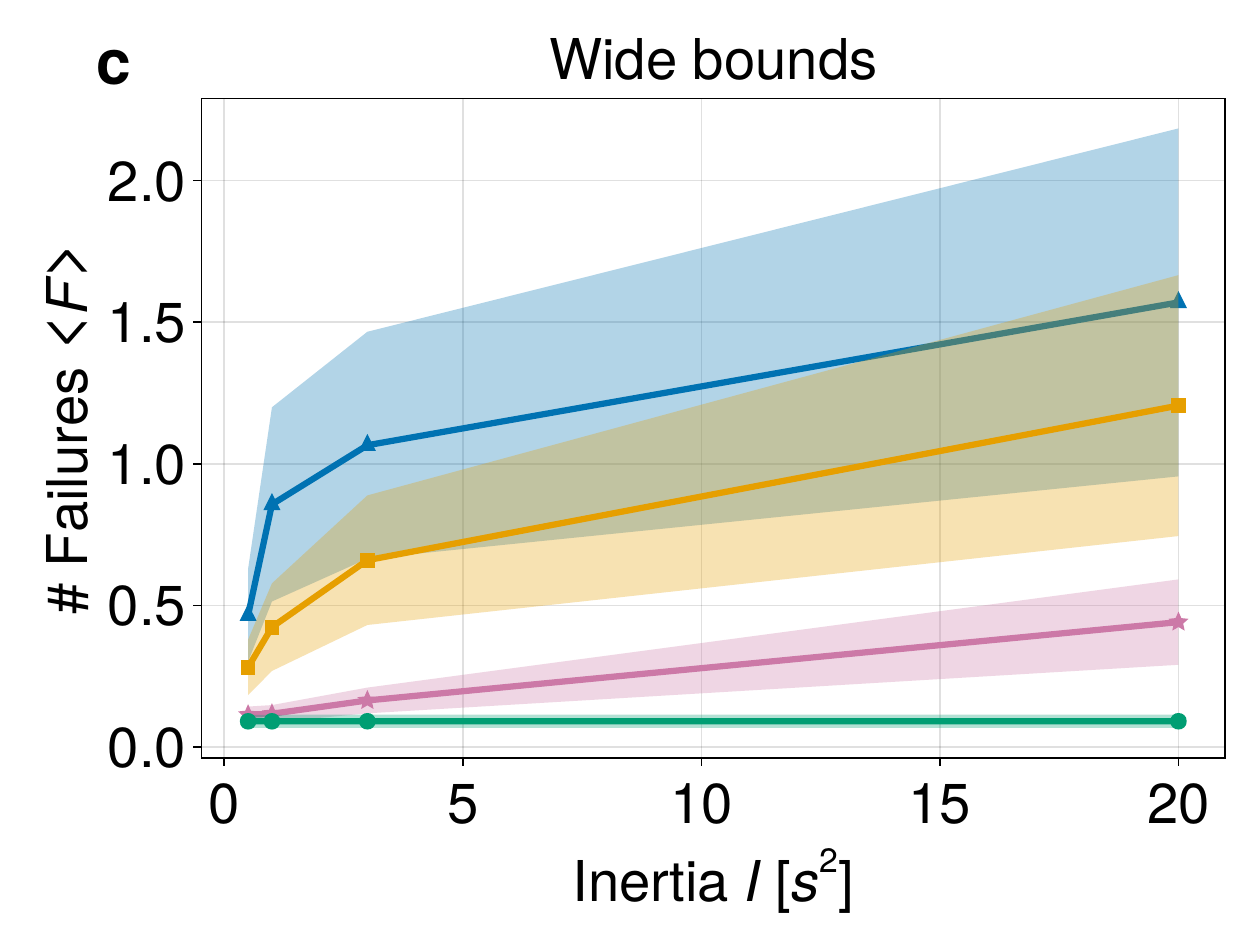}
  \caption{Number of failures $\langle F\rangle$ as a function of inertia $I$ for different fixed values of Damping $D$ for narrow bounds in \textbf{a}, intermediate bounds in \textbf{b} and wide bounds in \textbf{c}. 
  Watts-Strogatz networks are analyzed, and the full-failure model allows for both node and line failures. The standard error is shown in shaded bands. The number of failures $\langle F\rangle$ decreases with $D$, independent of the frequency bounds $f_b$.}
  \label{fig:WS_vary_D_only_uebergang_lines+nodes}
\end{figure}

In section~\ref{sec:inertia_effect}, three frequency bound regimes are analyzed: ``narrow'', ``intermediate'' and ``wide'' correspond to $f_b = 0.01, 0.03, 0.15\,\si{\hertz}$ for the WS networks and $f_b = 0.2, 0.3, 1.0\,\si{\hertz}$ for the power grid case. 
This is the case in Fig.~\ref{fig:inertia_effect} and in Supplementary Figs.~\ref{fig:WS_coupled_model_node_and_line_failures_separately},~\ref{fig:WS_uebergang_lines+nodes_different_scalings},~\ref{fig:WS_vary_D_only_uebergang_lines+nodes}.

\begin{figure}
  \includegraphics[width=0.5\linewidth]{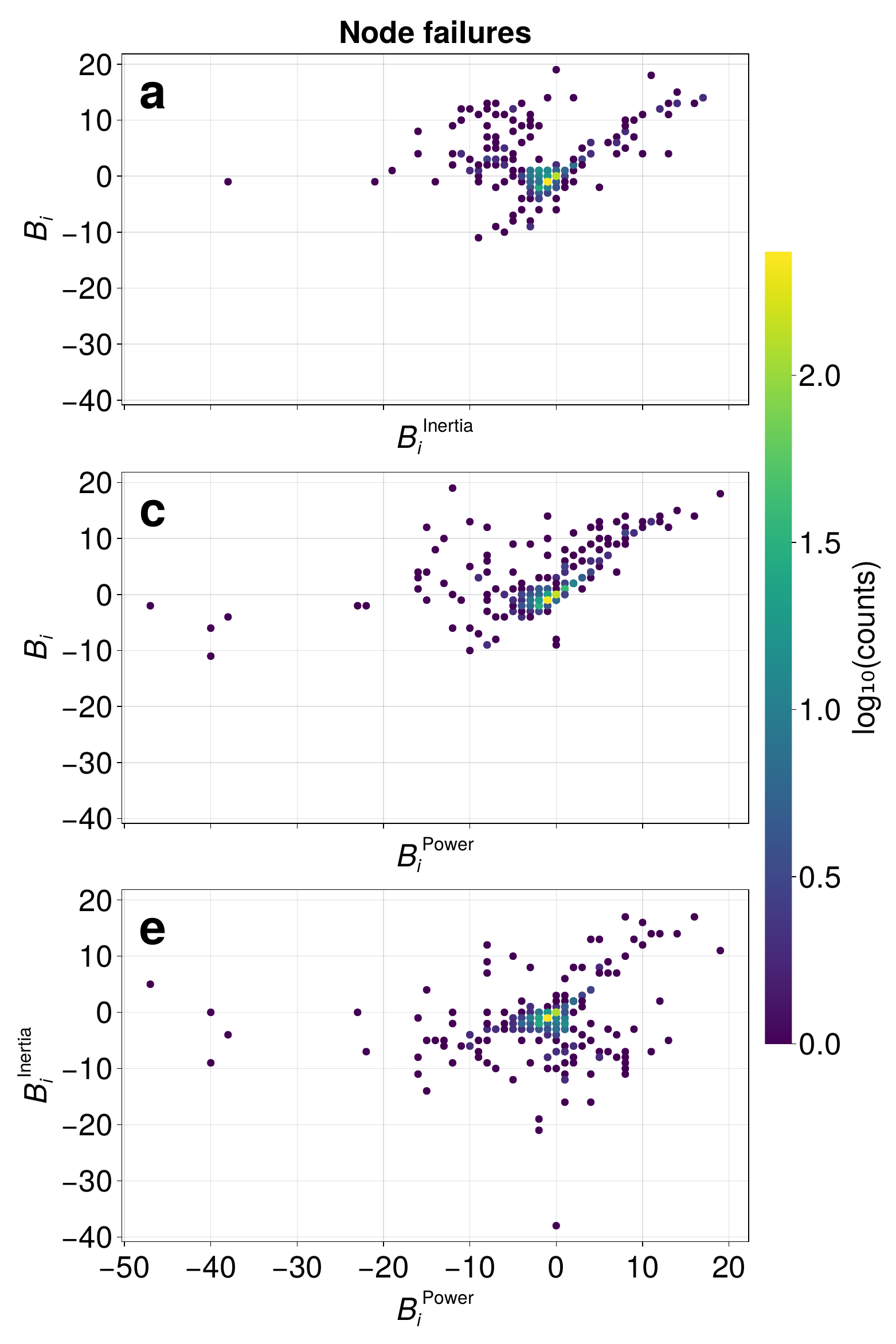}
  \includegraphics[width=0.5\linewidth]{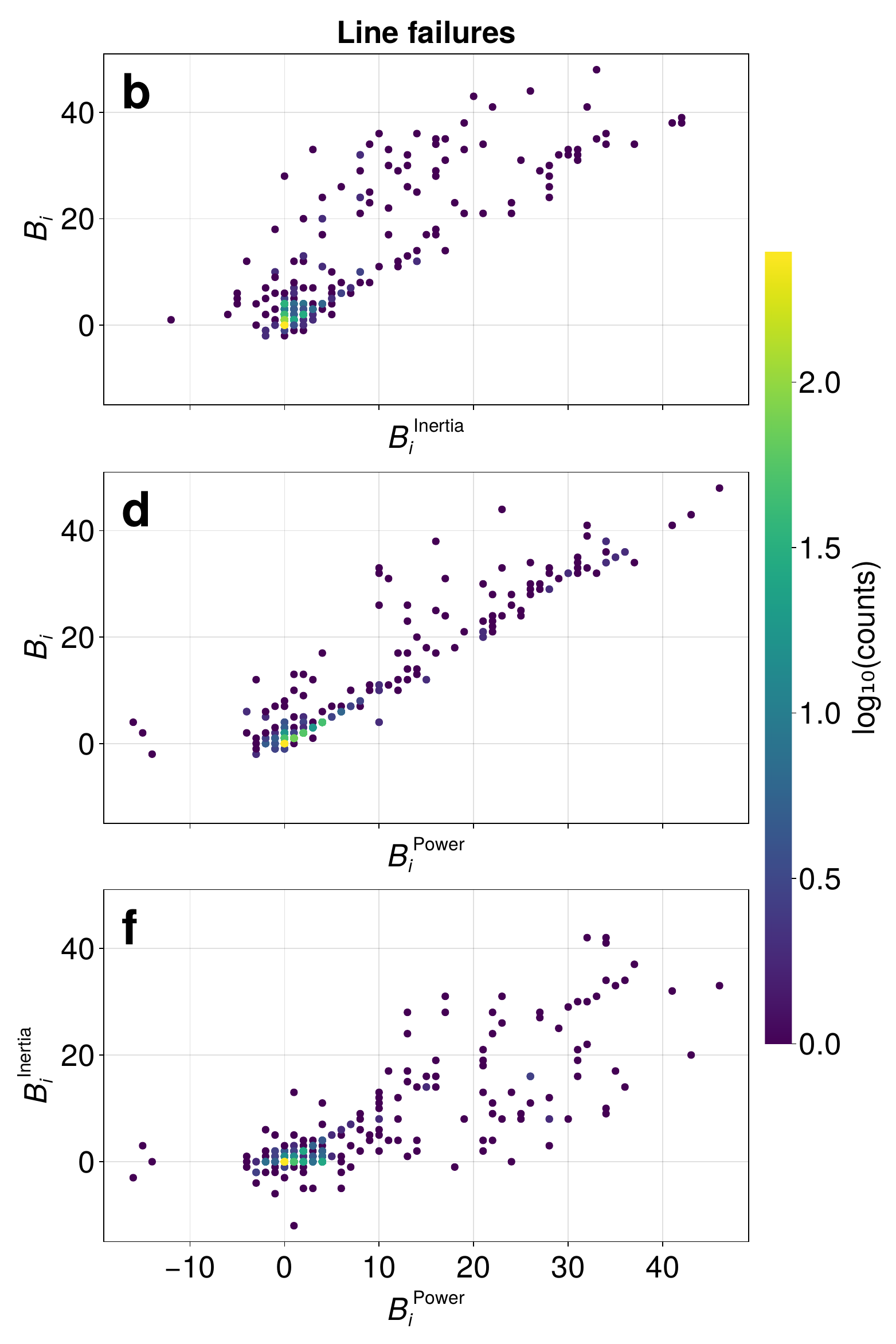}
  \caption{Braessness $B_i$ of individual cascades in the Watts–Strogatz ensemble for the same data as in Fig.~\ref{fig:WS_braessness_vs_braessness_2D_count_lines+nodes} for node failures in \textbf{a,c,e} and line failures in \textbf{b,d,f} separately. The node Braessness for triggering line $i$ is defined as $B_i = n^{\text{wide}}_i - n^{\text{narrow}}_i$. The line Braessness is defined as $B_i = l^{\text{wide}}_i - l^{\text{narrow}}_i$.}
  \label{fig:WS_braessness_vs_braessness_2D_count_lines_and_nodes_separately}
\end{figure}

In sections~\ref{sec:breaessness_effect} and~\ref{sec:mechanisms_braess}, we have only two frequency bounds. Here, ``narrow'' and ``wide'' correspond to $f_b=0.03,0.15\,\si{\hertz}$ with inertia $I=\SI{7.5}{\second\squared}$ for the WS ensemble and to $f_b = 0.3, 1.0\,\si{\hertz}$ with $I=\SI{3.0}{\second\squared}$ for the power grid case. This is the case in Figs. 
\ref{fig:WS_illustrative_fig},
\ref{fig:WS_braessness_histogramm_lines+nodes},
\ref{fig:WS_braessness_vs_braessness_2D_count_lines+nodes} 
and in Supplementary Figs. 
\ref{fig:histograms_braessness_line_node_failures_separately},
\ref{fig:WS_braessness_power_vs_inertia_vs_full_2D_lines+nodes},
\ref{fig:WS_braessness_vs_braessness_2D_count_lines_and_nodes_separately}.

\subsection{Conceptual analytic model} \label{sec:methods_analytic_model}
We derive the analytic expressions for the conceptual model in Figs.~\ref{fig:inertia_effect}e and~\ref{fig:inertia_effect}f.
In the two-node model, we consider a single swing equation node, see Eq. (\ref{eq:swing_equation_b}), connected by a line to a slack node~\cite{machowskiPowerSystemDynamics2020}. A slack node is a conceptual tool in power system modeling that can be interpreted as a large power plant or battery. By fixing its voltage phase angle to $\theta=0$, it serves as the reference point for all other nodes in the system and compensates for any power imbalance in the network. We further linearize the sinusoidal coupling in the swing equation, which yields a linear system that can be solved analytically. Setting the line reactance to $x=1$ for simplicity and dividing by $I$ then gives:
\begin{align}
    \dot\theta &= \omega, \\
    \dot{\omega} &= \frac{P}{I} - \frac{D}{I}\omega - \frac{1}{I}\theta.
\end{align}
This second-order system corresponds to a damped harmonic oscillator and can be written as:
\begin{equation}
    \ddot{\theta} + \frac{D}{I}\dot{\theta} + \frac{1}{I}\theta = \frac{P}{I}.
\end{equation}
Starting from the steady state, we perturb this system by instantaneously perturbing the power injection at the swing equation node by: $\Delta P = P'-P$ ($\Delta P=0.93$~p.u. in Figs.~\ref{fig:inertia_effect}e and~\ref{fig:inertia_effect}f). This perturbation induces a dynamic response in both the phase angle $\theta$ and the angular frequency $\omega$.
The homogeneous part of the equation determines the transient dynamics. The characteristic equation yields the eigenvalues:
\begin{equation}
    \lambda_{\pm} = \frac{-D \pm \sqrt{D^2 - 4I}}{2I}.
\end{equation}
With our chosen parameters, $\lambda_{\pm}$ are complex, implying an oscillatory response, i.e.:
\begin{equation} \label{eq:condition_complex_EV}
    D^2 < 4I.
\end{equation}
Defining:
\begin{align}
    \rho = \frac{P-P'}{\sin\left(\arctan\left(\frac{\Omega}{\nu}\right)\right)}, \qquad \phi & = \arctan\left(\frac{\Omega}{\nu} \right)
\end{align}
and using: 
\begin{equation}
    \nu = \frac{D}{2I}, \qquad \Omega = \frac{\sqrt{4I-D^2}}{2I},
\end{equation}
the system response after the perturbation becomes:
\begin{align}
    \label{eq:phase_slack_to_gen}
    \theta(t) &= \rho e^{-\nu t}\sin\left(\Omega t + \phi \right)+P',
    \\
    \label{eq:frequency_slack_to_gen}
    \omega(t) &= -\frac{\rho}{\sqrt{I}}\;e^{-\nu t}\sin(\Omega t)
\end{align}
The maximum phase deviation after the perturbation is:
\begin{equation}
    \theta^{\text{max}} = \Delta P \left[ 1 + \exp\!\left( -\frac{\pi}{\sqrt{\eta - 1}} \right) \right],
    \ \eta = \frac{4 I}{D^2}.
\end{equation}

With Eqs. (\ref{eq:ohms_law}) and (\ref{eq:apparent_power_flow}),
the maximum apparent power flow on the line \eqref{eq:apparent_power_sin} is then:
\begin{equation}
    |S^{\max}| = 2 \left| \sin\!\left( \frac{\theta^{\max}}{2} \right) \right|.
\end{equation}
The maximum absolute frequency deviation is given in Eq. (\ref{eq:f^max}).
Both analytic expressions describe the first maximum of the system’s transient response.
In addition to the linearization of Eq. (\ref{eq:swing_equation_b}), the model only considers the largest deviation in $|S|$ and $f$, which, due to the damped oscillation, always occurs at the first maximum. 
Therefore, $|S^{\text{max}}|$ and $|f^{\text{max}}|$ only capture the system’s initial response to the perturbation.

\section{Code and data availability}
All code and data to reproduce the findings are available at \doi{10.5281/zenodo.20341268}.

\section{Acknowledgements}
N.B. gratefully acknowledges the support from the Reiner Lemoine Foundation through a scholarship. The authors thank Philipp Blechinger for valuable comments and feedback on the manuscript, Edmund Obermeyer for careful proofreading, and Mehrnaz Anvari for fruitful discussions.
This project has received funding from the German Federal Ministry for Economic Affairs and Energy (BMWE) as part of the eKI4DS Project 03EI1092A and the OpPoDyn Project 03EI1071A.
The authors gratefully acknowledge the Ministry of Research, Science and Culture (MWFK) of Land Brandenburg for supporting this project by providing resources on the high performance computer system at the Potsdam Institute for Climate Impact Research.

\section{Author contributions}
All authors designed the failure model and the research. H.W. provided the initial codebase. N.B. extended the code and performed all simulations, data analyses, and figure preparation. A.B., F.H., and N.B. discussed and interpreted the results. N.B. and F.H. wrote the manuscript. All authors reviewed and edited the manuscript. F.H. and A.B. supervised the project.

\section{Competing interests}
The authors declare no competing interests.

\section{Additional information}
Correspondence should be addressed to Nubius Brandner or Frank Hellmann.
 \bibliography{NLCp}
\clearpage

\clearpage
\onecolumngrid  %
\beginsupplement

\begin{center} %
  {\LARGE Supplementary Information\par}
  \vspace{1.5ex}
  {\Large \papertitle\par}
  \vspace{1ex}
  {\large \paperauthors\par}
  \vspace{1ex}
  {\small \paperaffiliations}
\end{center}
\vspace{2em}

This Supplementary Information follows the general narrative of the main manuscript, adding to it more detailed analyses and parameter studies.
In particular, Supplementary section~\ref{sec:SI_inertia_variation} covers an example of a Watts-Strogatz network that illustrates how different inertia leads to different cascading behavior. It further covers how the variation of inertia affects the number of node and line failures individually. 
Supplementary section~\ref{sec:SI_damping_variation} includes the variation of Damping $D$.
Supplementary section~\ref{sec:SI_braess_section} shows the dynamical Braess's paradox in a broader parameter regime than presented in the main text. 
Supplementary section~\ref{sec:SI_mechanisms_braess} adds analyses of the power grid.
Furthermore, in Supplementary section~\ref{sec:SI_alpha_variation}, we present a variation of the safety parameter for the line robustness $\alpha$.
Finally, in Supplementary section~\ref{sec:SI_WS-networks}, we cover the effect of different rewiring probabilities $\beta$ for the Watts-Strogatz networks.

\section{Supplementary figures: The impact of power system inertia} \label{sec:SI_inertia_variation}

\begin{figure}
  \center
  \includegraphics[width=\textwidth]{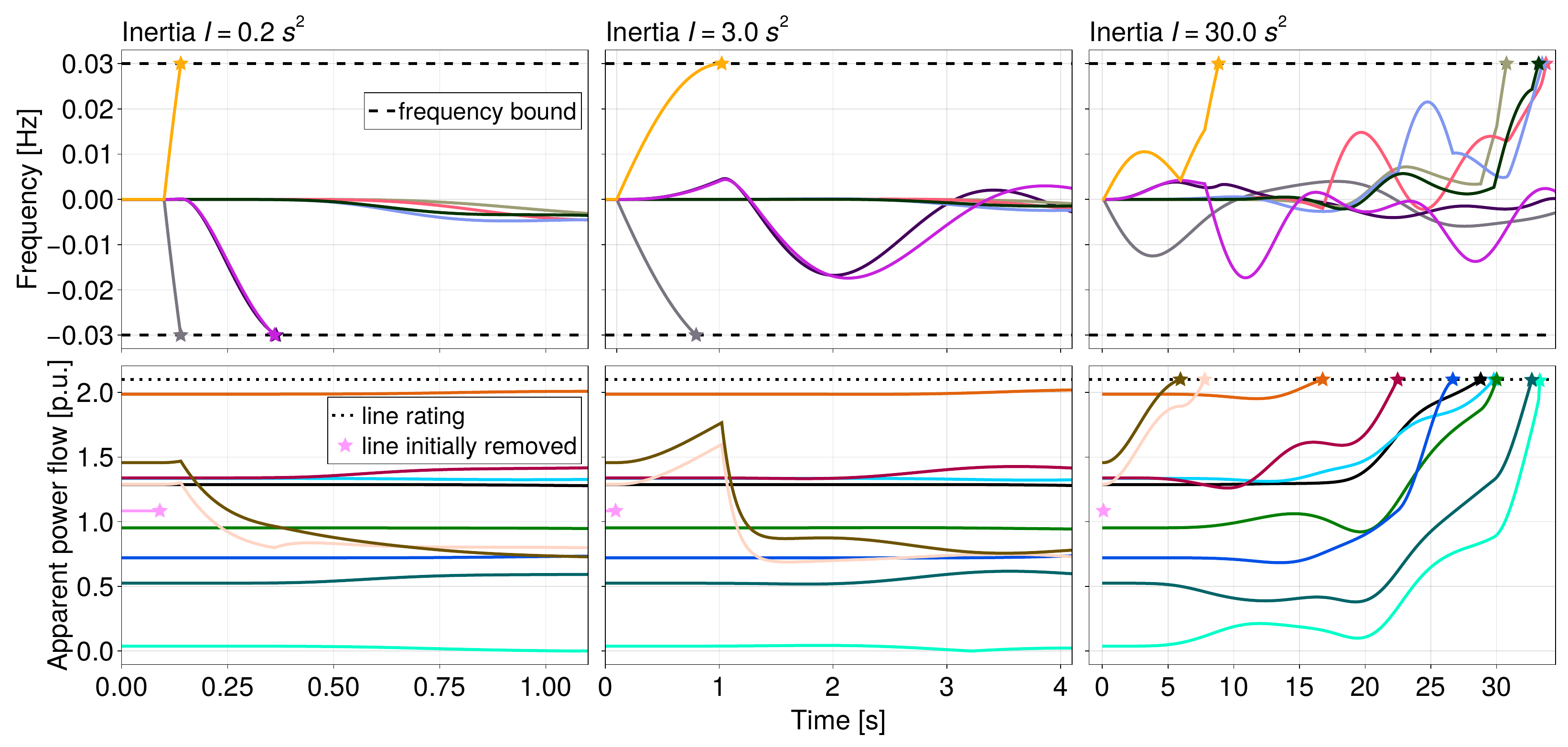}
  \caption{Frequency (top) and apparent power flow trajectories (bottom) for different inertia values for a network from the Watts-Strogatz ensemble with frequency bounds $f_b = \SI{0.03}{\hertz}$, where the full-failure model allows for both node and line failures.
  To trigger a cascade, the pink line is removed at $t = \SI{0.1}{\second}$.
  For three different inertia values $I = 0.2,\,3.0,\,30.0\,\si{\second\squared}$,
  we show the trajectories of same nodes (top) and lines (bottom), i.e., those that fail (indicated by stars) for at least one of the three inertia values.
  For $I = \SI{0.2}{\second\squared}$, four nodes fail, while for $I = \SI{3.0}{\second\squared}$, only two fail; no lines fail in either case. In contrast, for $I = \SI{30.0}{\second\squared}$, multiple lines fail, which potentially leads to multiple node failures.
  This example illustrates how varying inertia $I$ leads to markedly different trajectories and cascading dynamics within the same network. It highlights the emergence of genuine cascading behavior with secondary failures, as opposed to scenarios where all failures occur
  shortly after the perturbation. Moreover, it demonstrates the interplay between line and node failures.
  }  
  \label{fig:WS_trajectories_inertia_variation}
\end{figure}

\begin{figure}
  \center
  \includegraphics[width=.5\linewidth]{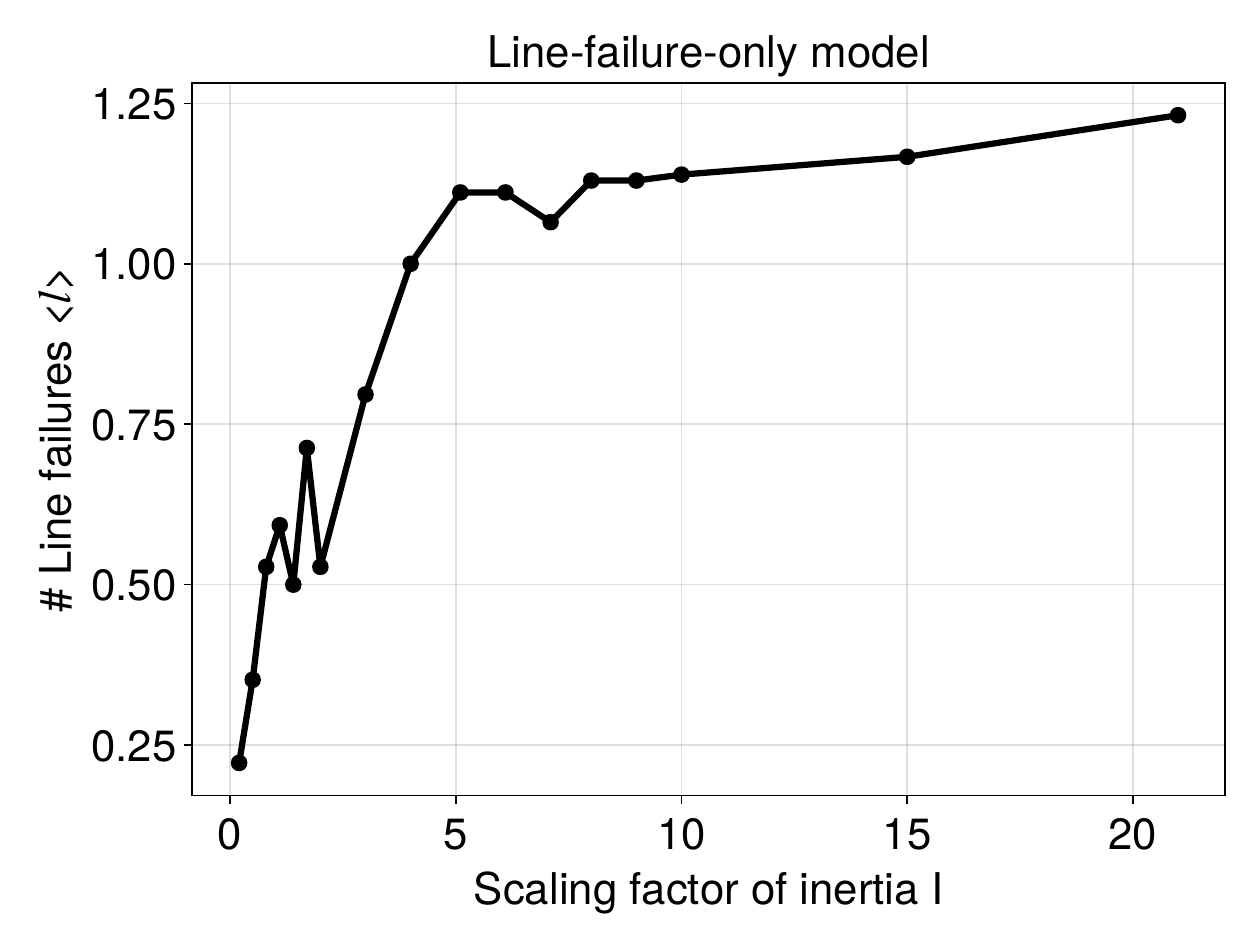}
  \caption{Number of line failures $\langle l \rangle$ as a function of inertia $I$ in the RTS-GMLC power grid test system~\cite{barrowsIEEEReliabilityTest2020} for the line-failure-only model, where the decoupled line failure model exclusively allows for line failures. The inertia parameter at the nodes is not heterogeneous and is thus scaled by a homogeneous factor. As the scaling factor of inertia increases, the number of line failures increases.}
  \label{fig:line_failures_test_case}
\end{figure}

\begin{figure}
  \center
  \includegraphics[width=0.5\linewidth]{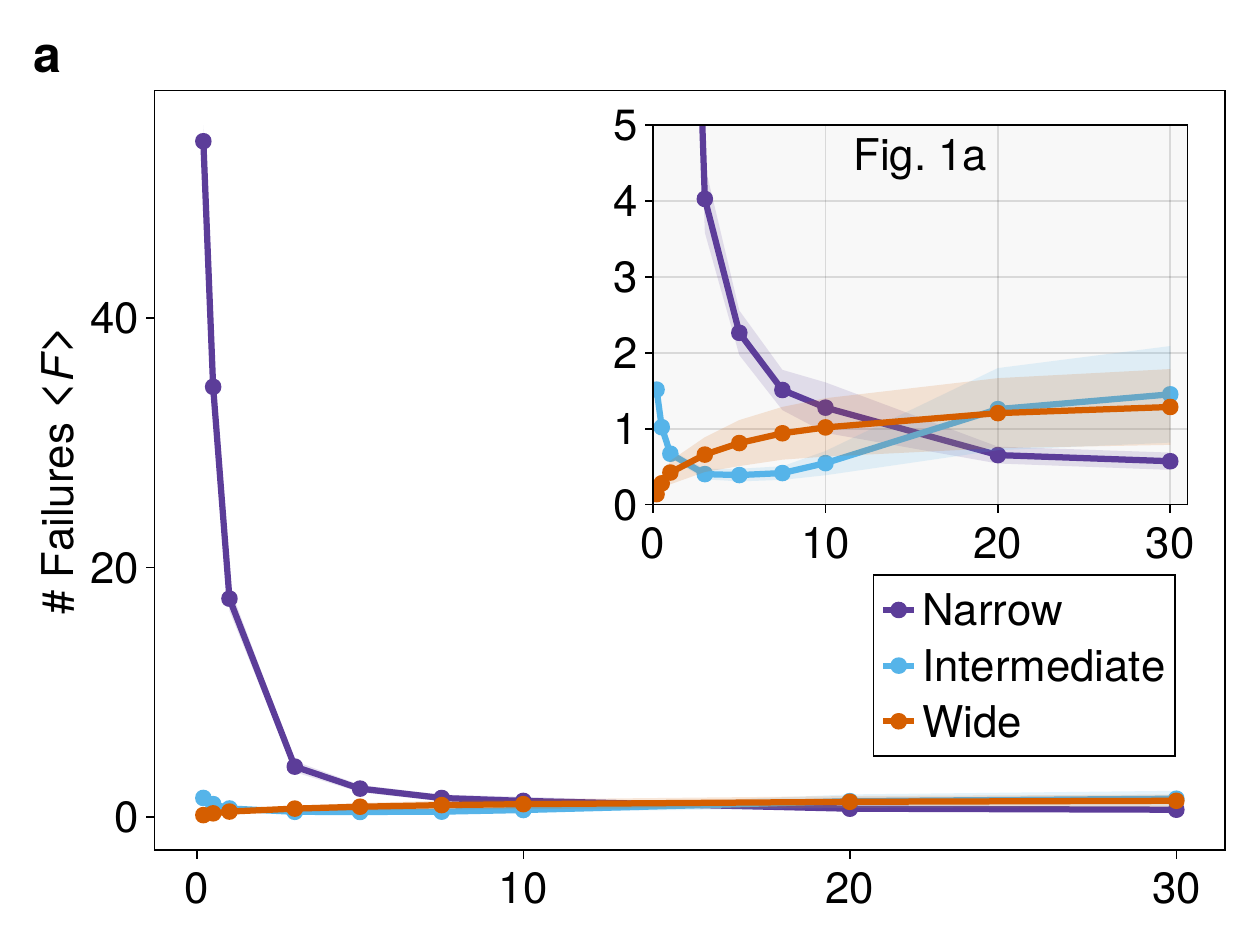}
  \\ 
  \includegraphics[width=.5\linewidth]{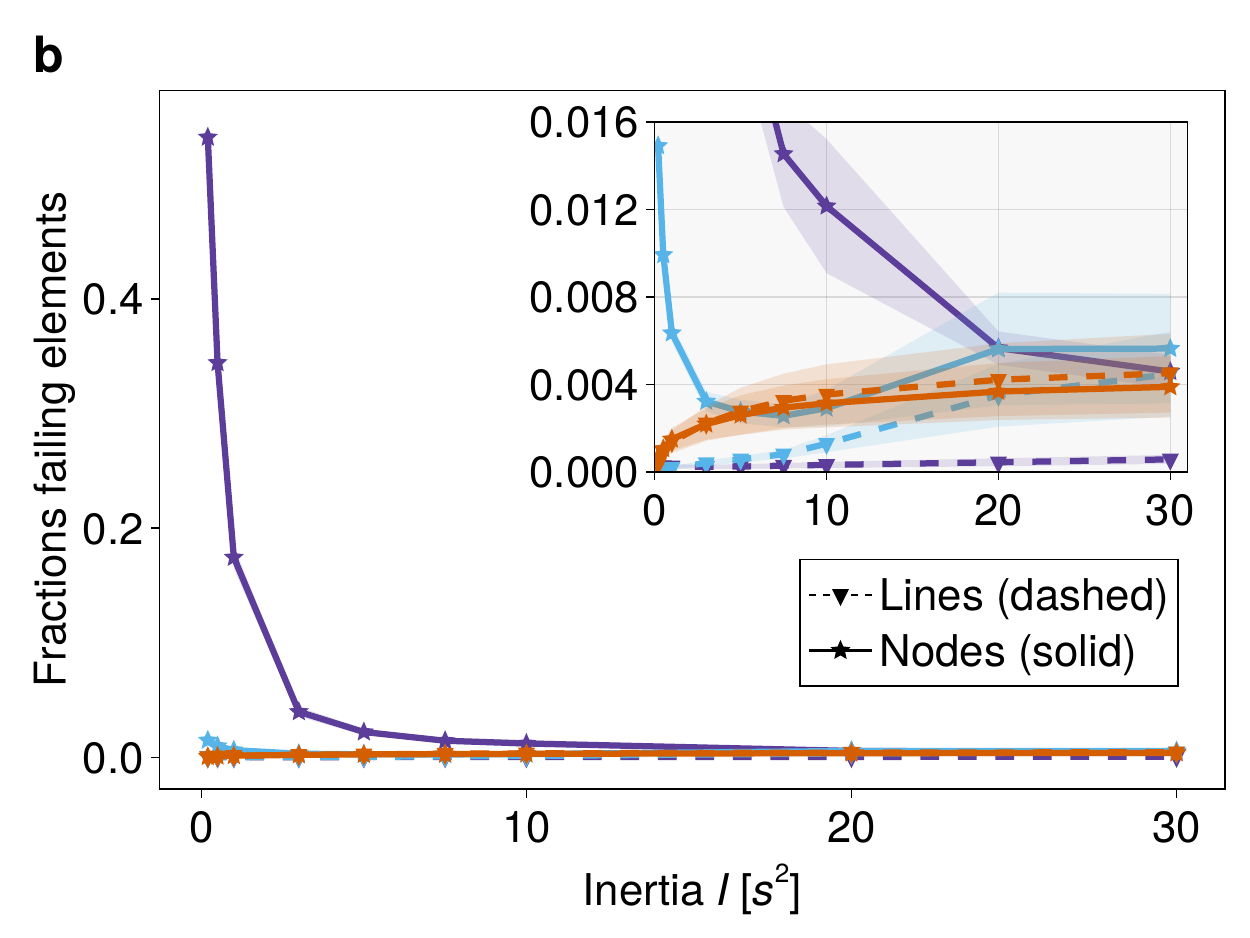}
  \caption{Failures as a function of inertia $I$ for different frequency bounds. Watts-Strogatz networks are analyzed and the full-failure model allows for both node and line failures. The standard error is shown in shaded bands.
  Panel \textbf{a} shows the number of failures $\langle F\rangle$ for low inertia values at narrow bounds, while the inset indicates the region in Fig.~\ref{fig:inertia_effect}a.
  In \textbf{b}, the fractions of node failures $\frac{\langle n\rangle}{N}$ (solid) and line failures $\frac{\langle l\rangle}{M}$ (dashed) are shown separately. For narrow bounds, line failures are almost absent for all inertia values, while node failures follow the behavior of the node-failure-only model. For wide bounds, line failures and node failures both increase with inertia, with the same qualitative behavior as in the line-failure-only model, indicating that the observed node failures are caused by failures of adjacent lines. For intermediate bounds, node failures first decrease, while line failures increase as inertia is ramped up. Eventually, node failures start rising again, proportional to the increase in line failures. There is a trade-off between inertia preventing node failures and increasing line failures, and an optimal level of inertia exists.}
  \label{fig:WS_coupled_model_node_and_line_failures_separately}
\end{figure}

\newpage
\section{Supplementary Figures: Variation of the damping \texorpdfstring{$D$}{D}} \label{sec:SI_damping_variation}
In this Supplementary section, we vary the damping parameter $D$. In all other sections (except Supplementary section~\ref{sec:SI_alpha_variation}), we set $D = \SI{1}{\second}$.

\section{Supplementary figures: The impact of nodal robustness} \label{sec:SI_braess_section}

\begin{figure}
  \includegraphics[width=0.5\linewidth]{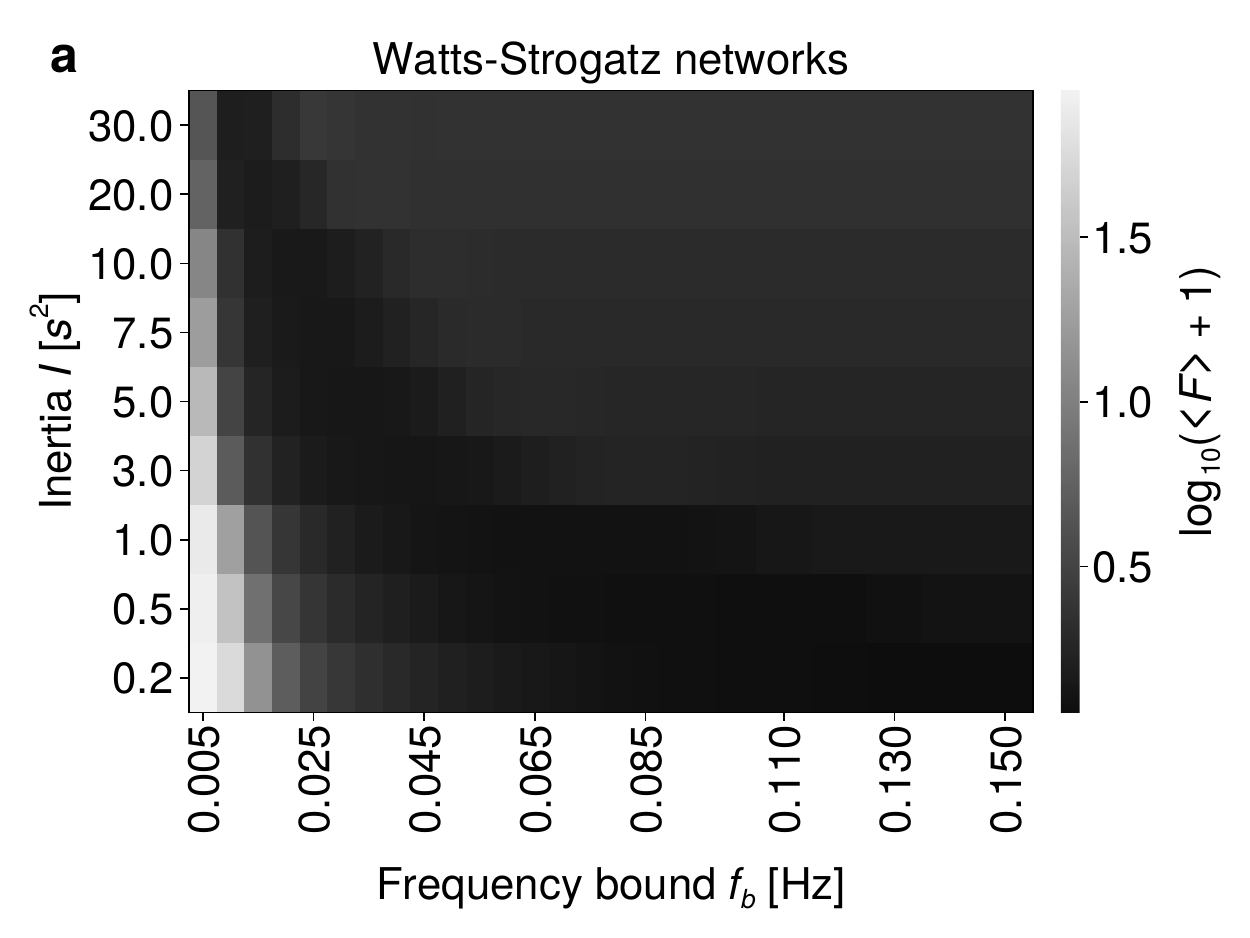}
  \includegraphics[width=0.5\linewidth]{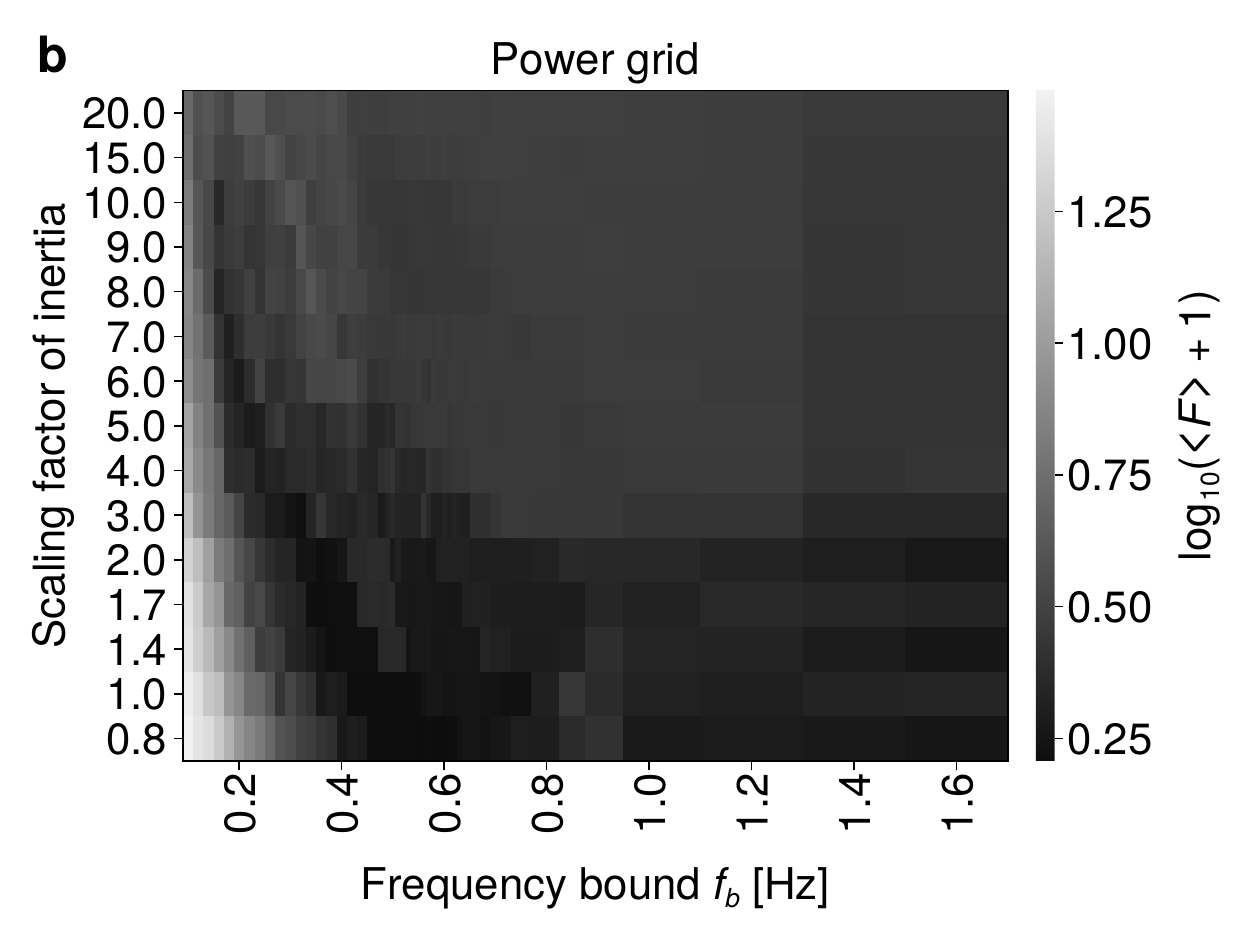}
  \caption{Number of failures $\langle F\rangle$ for a range of frequency bounds $f_b$ and different inertia values, where the full-failure model allows for both node and line failures. In \textbf{a}, the Watts-Strogatz networks and in \textbf{b}, the RTS-GMLC power grid test system~\cite{barrowsIEEEReliabilityTest2020} are analyzed. Note that the y-axes are nonlinearly scaled.
  We observe that increasing the frequency bounds $f_b$, and thus enhancing nodal robustness, can paradoxically increase the network's vulnerability to cascades, indicating a dynamical analog of Braess's paradox. This is highly robust across all inertia values. For $I>\SI{0.5}{\second\squared}$ in \textbf{a}, and across all inertia scalings in \textbf{b}, we observe a maximum in $\langle F\rangle$ at an intermediate $f_b$.}
  \label{fig:heatmap_f_b_vs_inertia}
\end{figure}

\begin{figure}
  \center
  \includegraphics[width=0.65\linewidth]{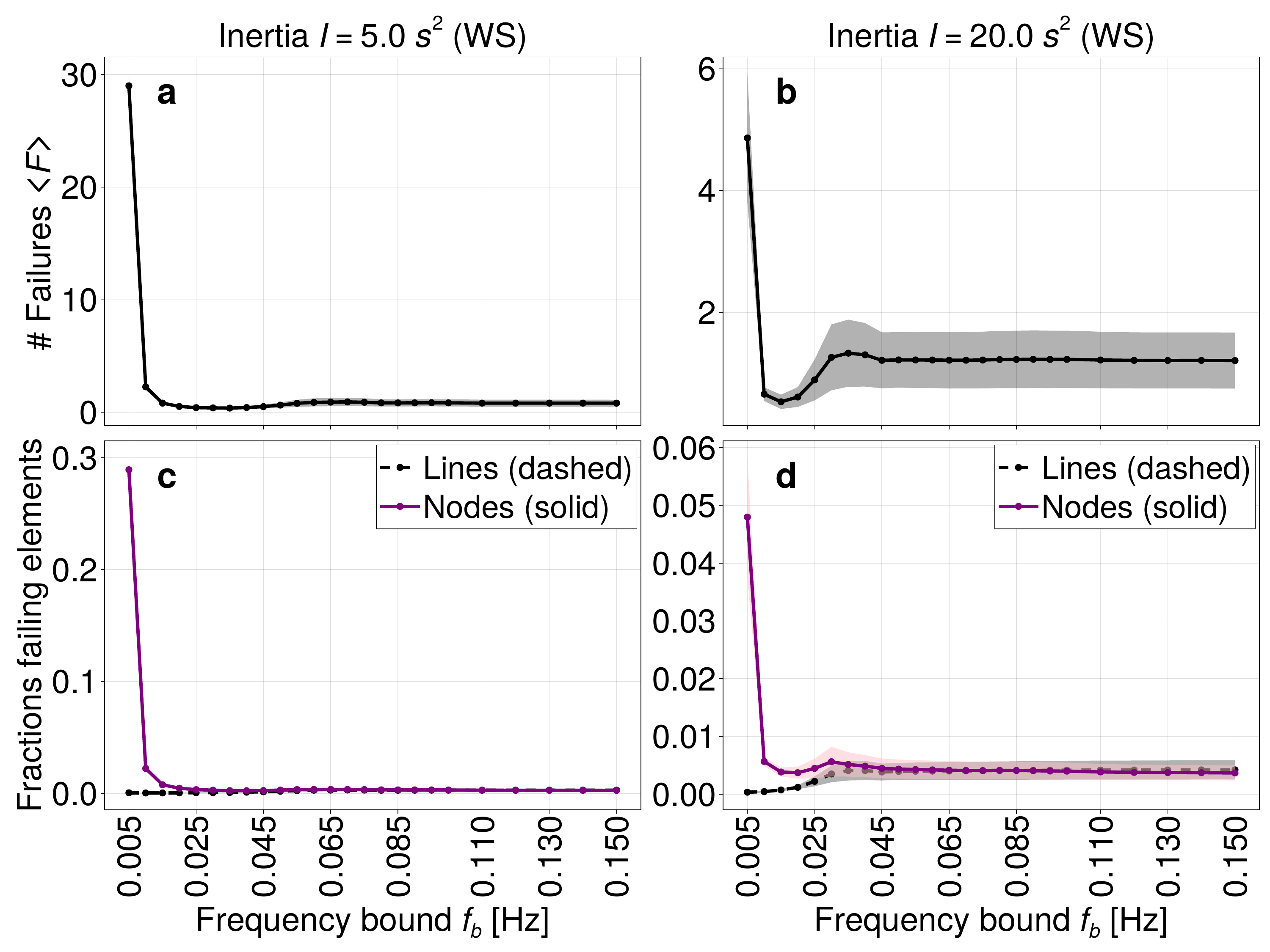}
  \caption{Number of failures as a function of the frequency bounds $f_b$ of nodes for the Watts-Strogatz networks with inertia $I=\SI{5.0}{\second\squared}$ in \textbf{a,c} and $I=\SI{20.0}{\second\squared}$ in \textbf{b,d}, where the full-failure model allows for both node and line failures. Panels \textbf{a,b} show the summed average of all failures $\langle F\rangle$, while \textbf{c,d} show the fractions of node failures $\frac{\langle n\rangle}{N}$ and line failures $\frac{\langle l\rangle}{M}$ separately. The standard error is shown in shaded bands. In \textbf{c}, for sufficiently high inertia, widening the frequency bounds $f_b$ can paradoxically increase the number of failures, revealing a local maximum and indicating a dynamical analog of Braess's paradox. The same qualitative behavior appears when the normalized average number of node failures $\frac{\langle n\rangle}{N}$ and line failures $\frac{\langle n\rangle}{M}$ are considered separately in \textbf{d}. In \textbf{a} and \textbf{b}, with significantly lower inertia, we do not see the effect.}
  \label{fig:WS_f_b_vs_failures}
\end{figure}

\begin{figure}
  \center
  \includegraphics[width=0.65\linewidth]{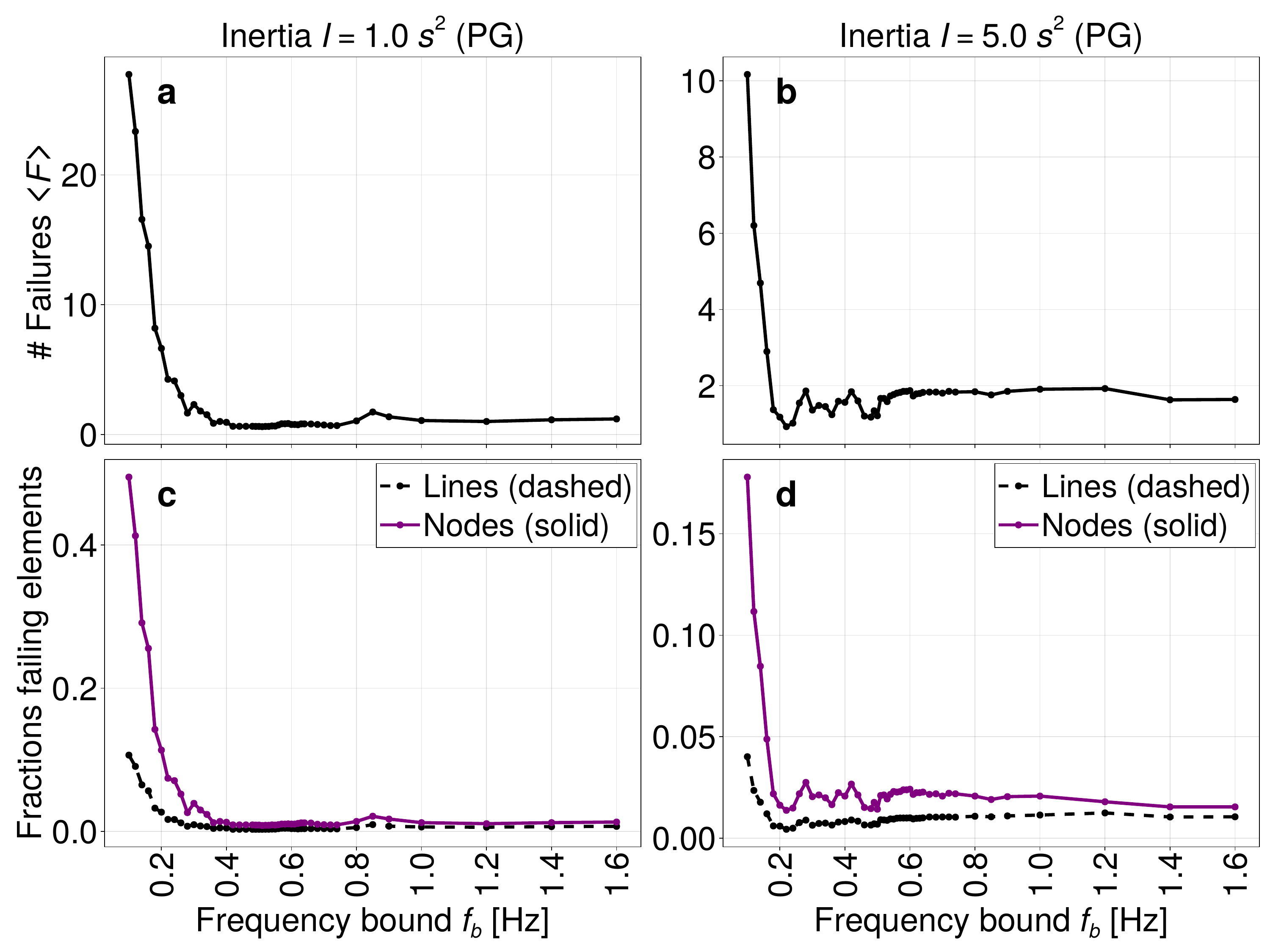}
  \caption{Number of failures as a function of the frequency bounds $f_b$ of nodes for the RTS-GMLC power grid test system~\cite{barrowsIEEEReliabilityTest2020} with inertia $I=\SI{1.0}{\second\squared}$ in \textbf{a,c} and $I=\SI{5.0}{\second\squared}$ in \textbf{b,d}, where the full-failure model allows for both node and line failures. Panels \textbf{a,b} show the summed average of all failures $\langle F\rangle$, while \textbf{c,d} show the fractions of node failures $\frac{\langle n\rangle}{N}$ and line failures $\frac{\langle l\rangle}{M}$ separately. In \textbf{c}, for sufficiently high inertia, widening the frequency bounds $f_b$ can paradoxically increase the number of failures, revealing a local maximum and indicating a dynamical analog of Braess's paradox. The same qualitative behavior appears when the normalized average number of node failures $\frac{\langle n\rangle}{N}$ and line failures $\frac{\langle n\rangle}{M}$ are considered separately in \textbf{d}. In \textbf{a} and \textbf{b}, where there is less inertia, the effect is less pronounced.}
  \label{fig:RTS_f_b_vs_failures}
\end{figure}

\section{Supplementary Figures: Mechanisms underlying Braess's paradox} \label{sec:SI_mechanisms_braess}

\begin{figure}
  \includegraphics[width=0.5\linewidth]{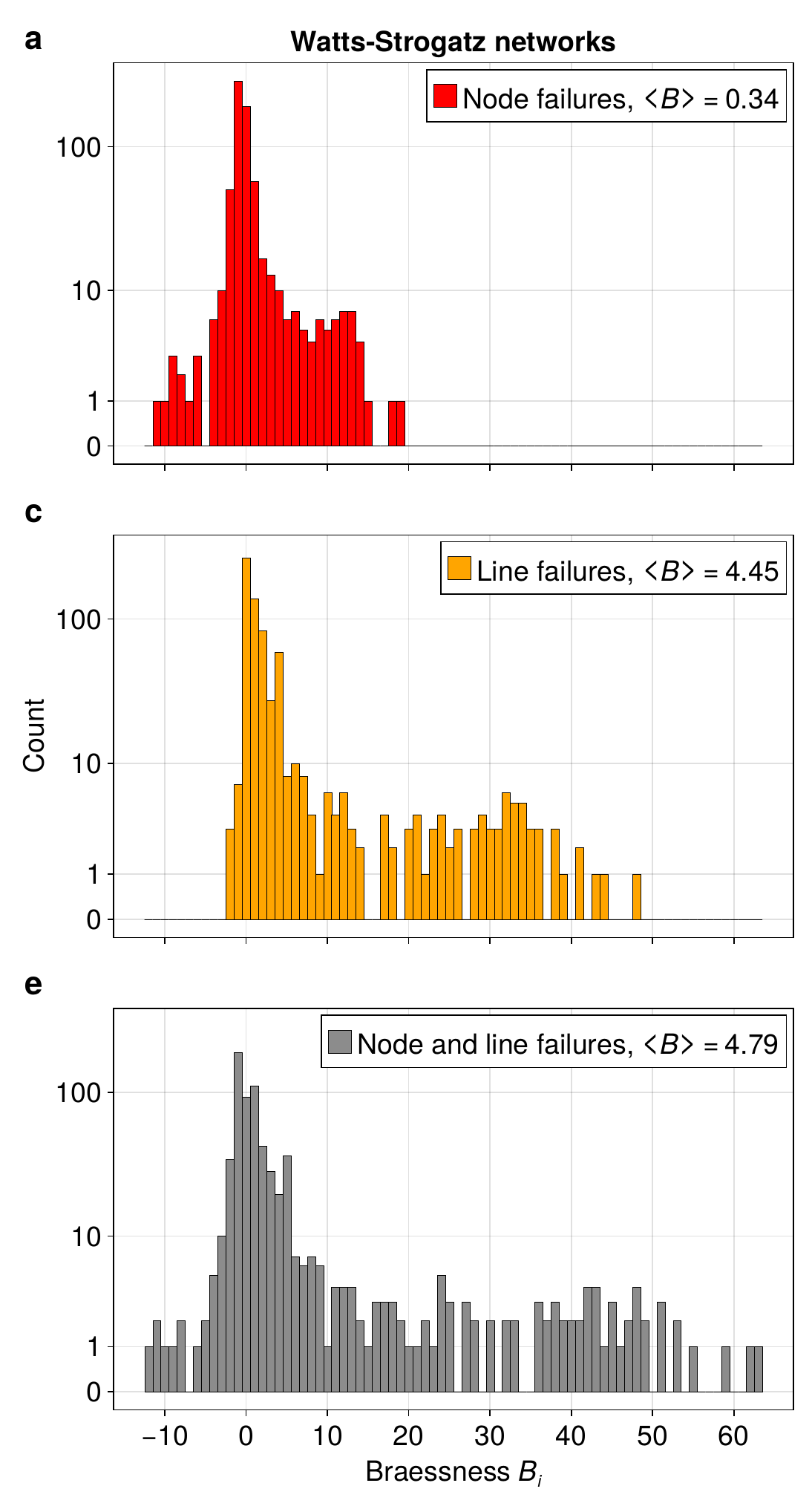}
\includegraphics[width=0.5\linewidth]{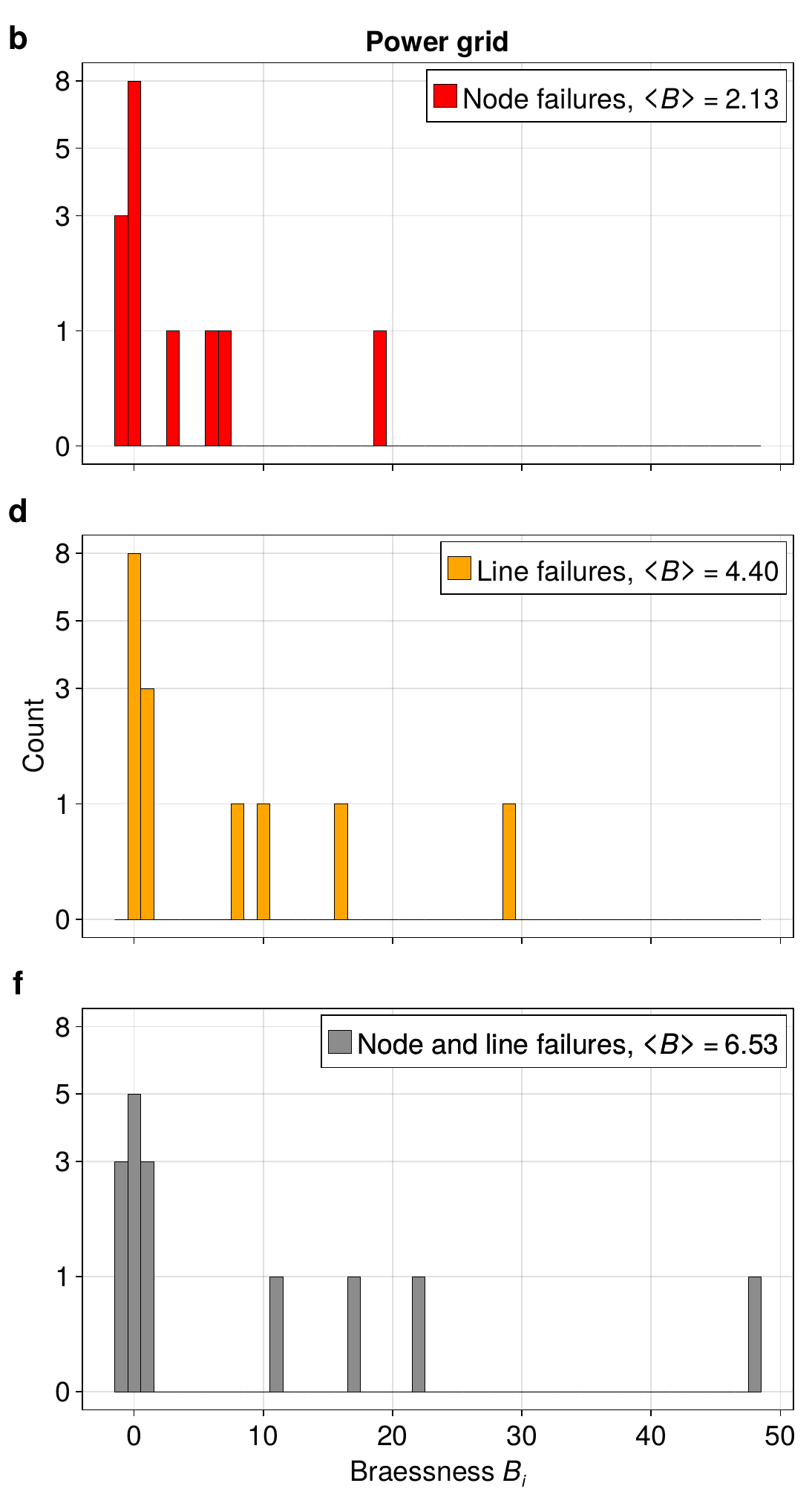}  
  \caption{Histogram of Braessness and average Braessness $\langle B\rangle$ for node failures, line failures and summed node and line failures separately. The full-failure model allows for both node and line failures. The Watts-Strogatz networks are analyzed in \textbf{a,c,e} on the left and the power grid in \textbf{b,d,f} are analyzed on the right.
  The node Braessness for triggering line $i$ is defined as $B_i = n^{\text{wide}}_i - n^{\text{narrow}}_i$. The line Braessness is defined as $B_i = l^{\text{wide}}_i - l^{\text{narrow}}_i$. Note that failures are not normalized with $N=200$ nodes and $M=100$ lines for the Watts-Strogatz networks and $N=73$ nodes and $M=108$ lines for the power grid.}
  \label{fig:histograms_braessness_line_node_failures_separately}
\end{figure}

\section{Supplementary Figures: Variation of the line robustness \texorpdfstring{$\alpha$}{alpha}} \label{sec:SI_alpha_variation}

\section{Supplementary Figures: Watts-Strogatz networks} \label{sec:SI_WS-networks}

\begin{figure}
  \includegraphics[width=0.33\linewidth]{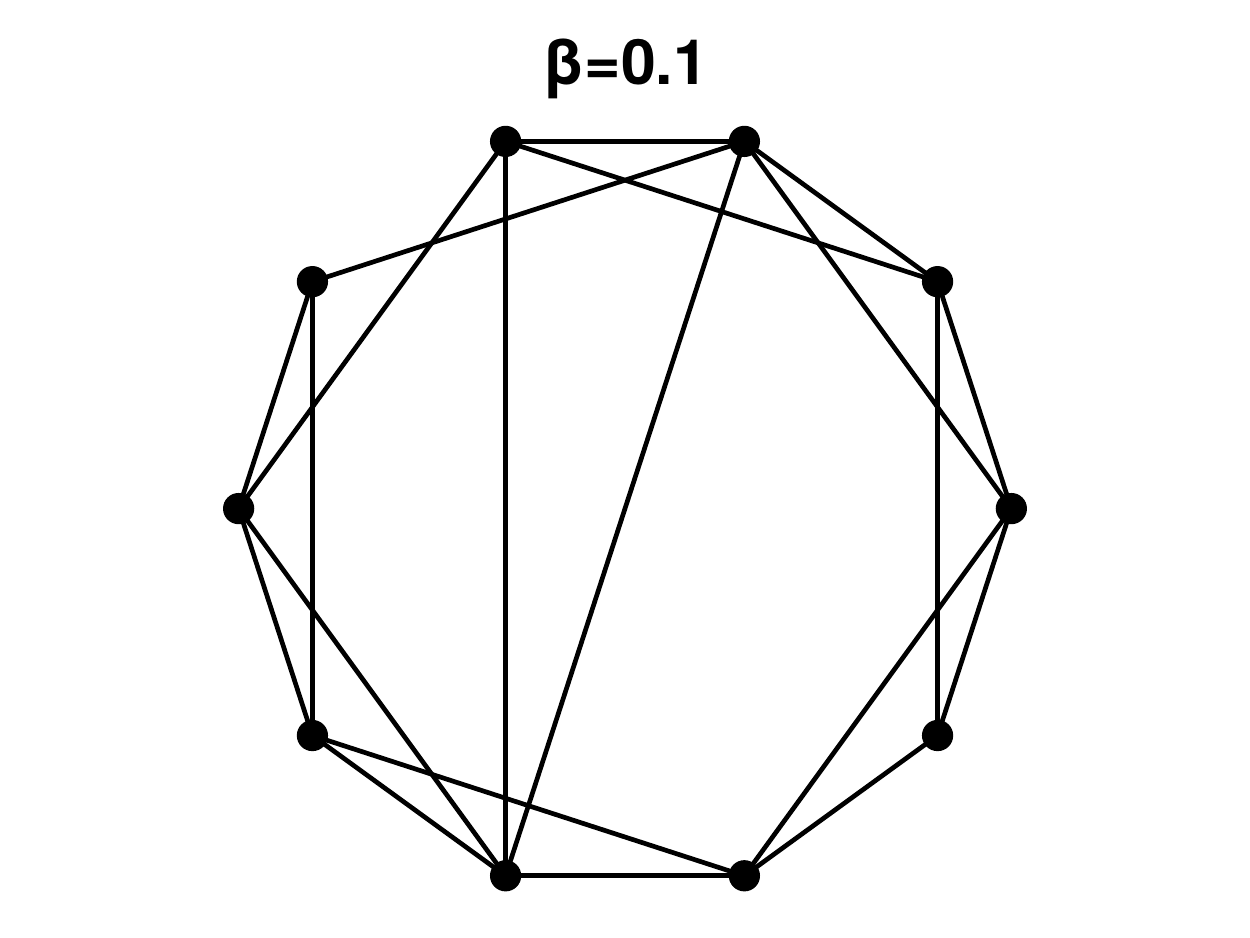}
  \includegraphics[width=0.33\linewidth]{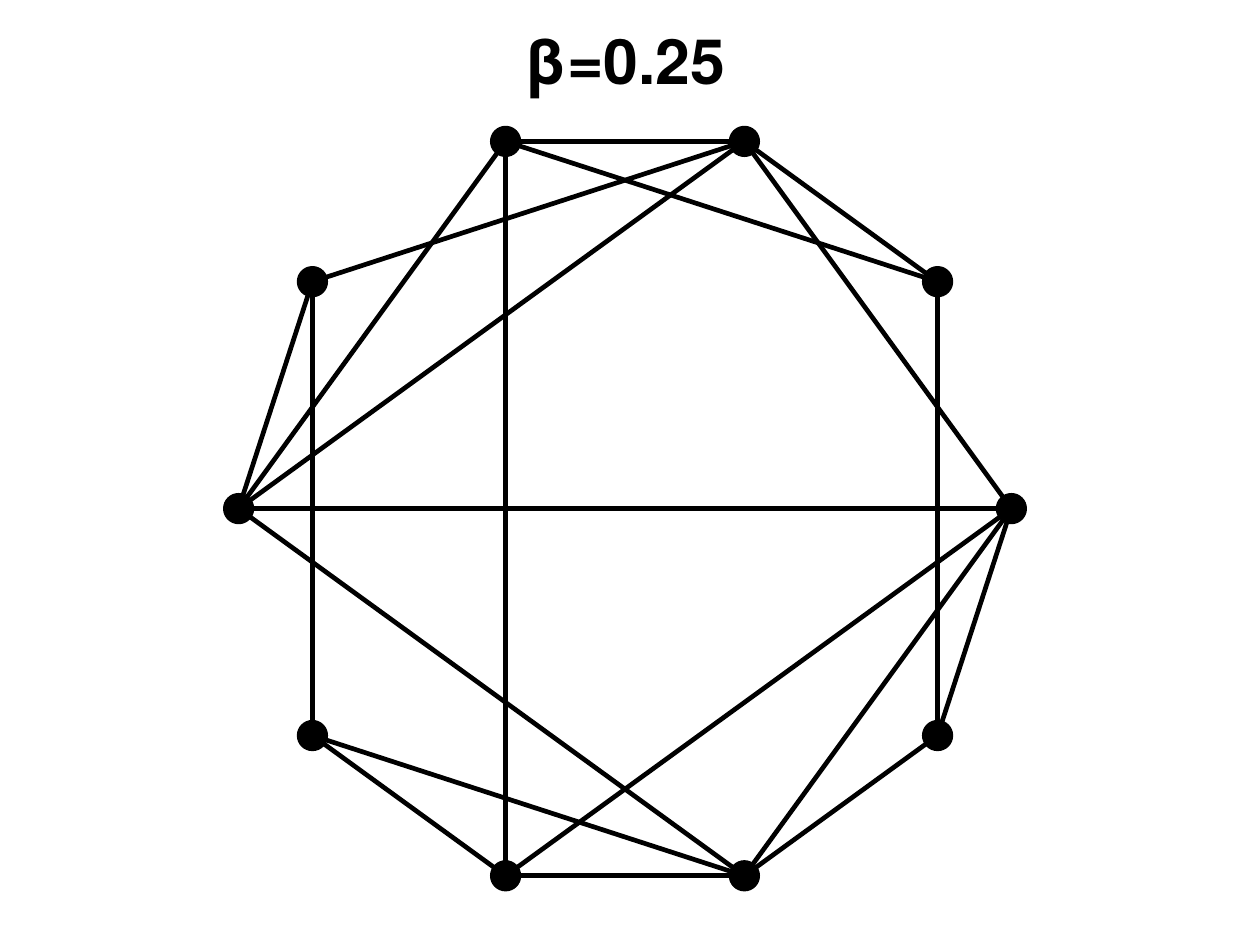}
  \includegraphics[width=0.33\linewidth]{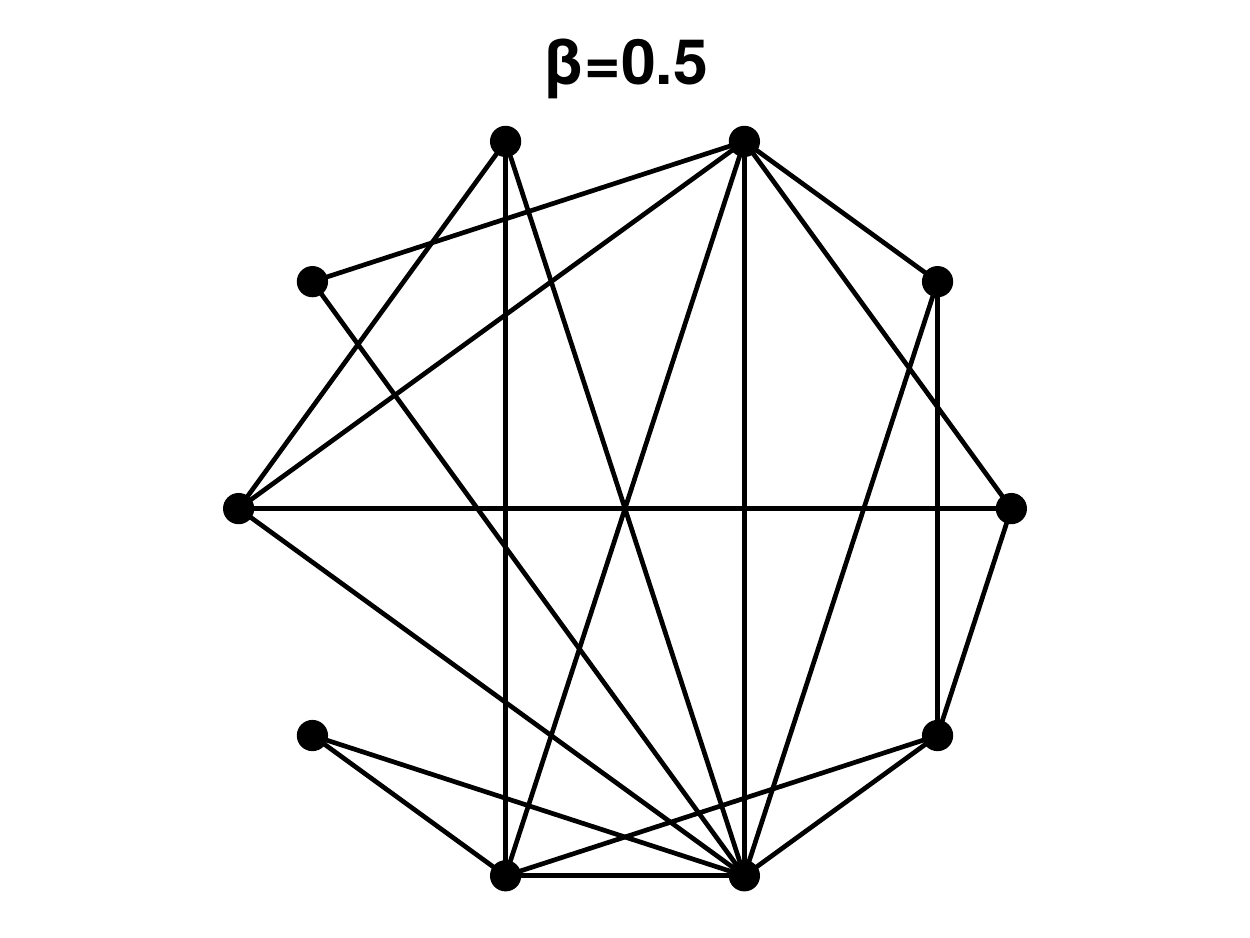}
  \caption{Watts-Strogatz networks with $N=10$ nodes, mean degree $k=4$ for different rewiring probabilities~$\beta$.}
  \label{fig:WS_example_topologies}
\end{figure}

\begin{figure}
  \center
  \includegraphics[width=0.6\textwidth]{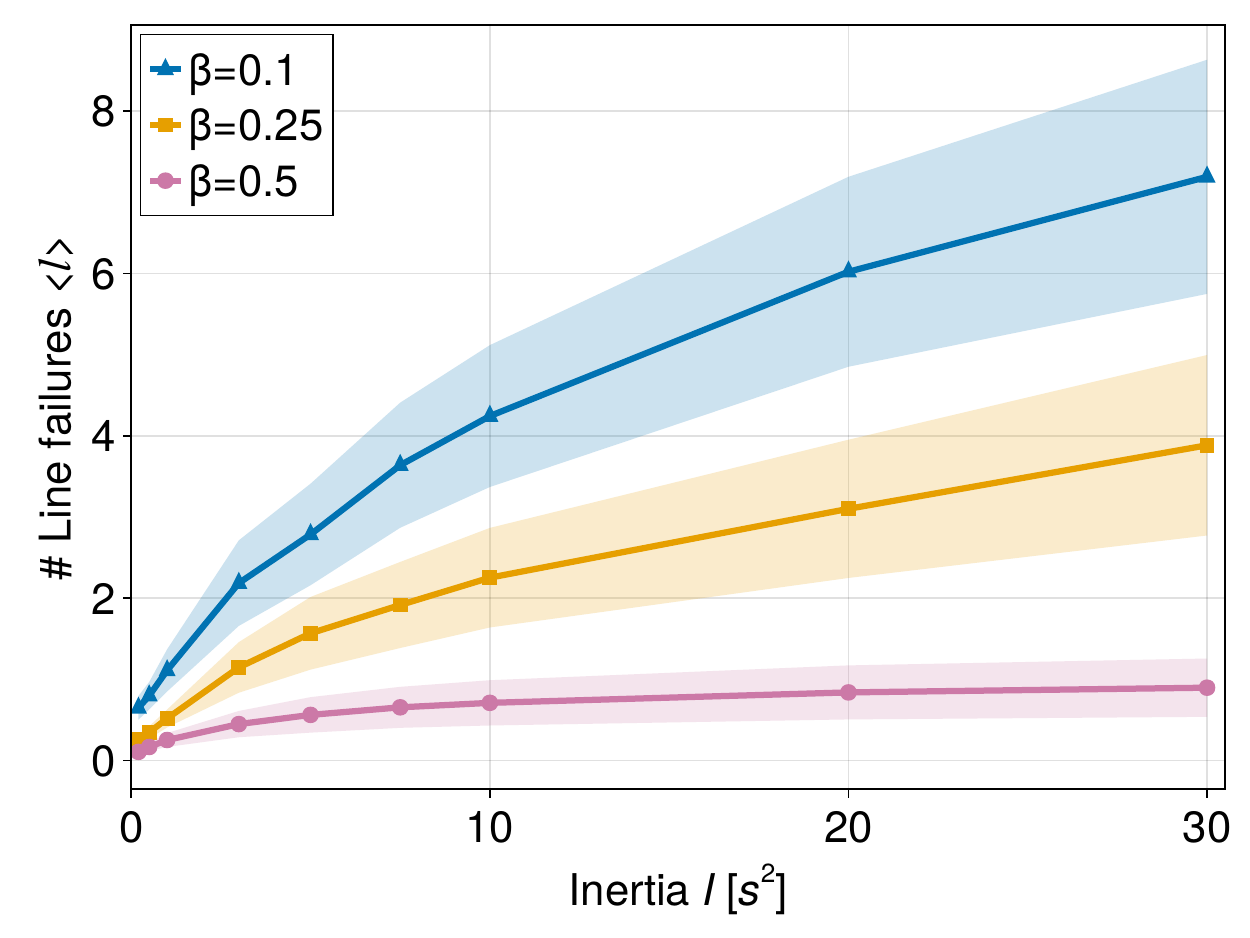}
  \caption{Average number of line failures $\langle l\rangle$ for the Watts-Strogatz networks across different rewiring probabilities $\beta$ for the line-failure-only model, where the decoupled line failure model exclusively allows for line failures. Shaded bands indicate the standard error. Larger $\beta$ lead to fewer line failures.}
  \label{fig:WS_lines_only_different_beta}
\end{figure}

\renewcommand{\bibsection}{\section*{Supplementary References}}
 
\end{document}